\pdfoutput=1
\documentclass[graybox]{SNmult}

\usepackage{newtxtext,newtxmath}
\usepackage{helvet}
\usepackage{courier}
\usepackage{type1cm}
\usepackage{makeidx}
\usepackage{graphicx}
\usepackage{multicol}
\usepackage[bottom]{footmisc}
\usepackage[numbers]{natbib}
\usepackage{url}
\usepackage{xurl} 
\usepackage{booktabs}

\usepackage{array}
\usepackage[T1]{fontenc}
\usepackage[utf8]{inputenc}
\usepackage[ngerman,english]{babel}

\makeatletter
\newcommand{\reprintbibliography}{%
  \begingroup\@fileswfalse\begin{thebibliography}{10}
\providecommand{\url}[1]{{#1}}
\providecommand{\urlprefix}{URL }
\expandafter\ifx\csname urlstyle\endcsname\relax
  \providecommand{\doi}[1]{DOI~\discretionary{}{}{}#1}\else
  \providecommand{\doi}{DOI~\discretionary{}{}{}\begingroup
  \urlstyle{rm}\Url}\fi

\bibitem{abrahao2025software}
Abrah{\~a}o, S., Grundy, J., Pezz{\`e}, M., Storey, M.A., Tamburri, D.A.:
  Software engineering by and for humans in an {AI} era.
\newblock ACM Transactions on Software Engineering and Methodology
  \textbf{34}(5) (2025).
\newblock \doi{10.1145/3715111}

\bibitem{agrawal2018prediction}
Agrawal, A., Gans, J., Goldfarb, A.: Prediction machines: The simple economics
  of artificial intelligence.
\newblock Harvard Business Press (2018)

\bibitem{ahmed2025artificial}
Ahmed, I., Aleti, A., Cai, H., Chatzigeorgiou, A., He, P., Hu, X., Pezz{\`e},
  M., Poshyvanyk, D., Xia, X.: Artificial intelligence for software
  engineering: The journey so far and the road ahead.
\newblock ACM Transactions on Software Engineering and Methodology
  \textbf{34}(5) (2025).
\newblock \doi{10.1145/3719006}

\bibitem{alenezi2026rethinking}
Alenezi, M.: Rethinking software engineering for agentic ai systems (2026).
\newblock \urlprefix\url{https://arxiv.org/abs/2604.10599}

\bibitem{armbrust2009above}
Armbrust, M., Fox, A., Griffith, R., Joseph, A.D., Katz, R.H., Konwinski, A.,
  Lee, G., Patterson, D.A., Rabkin, A., Stoica, I., et~al.: Above the clouds: A
  berkeley view of cloud computing.
\newblock Tech. rep., Technical Report UCB/EECS-2009-28, EECS Department,
  University of California~… (2009)

\bibitem{arthur1989competing}
Arthur, W.B.: Competing technologies, increasing returns, and lock-in by
  historical events.
\newblock The Economic Journal \textbf{99}(394), 116--131 (1989).
\newblock \doi{10.2307/2234208}

\bibitem{becker2025measuring}
Becker, J., Rush, N., Barnes, E., Rein, D.: Measuring the impact of early-2025
  ai on experienced open-source developer productivity (2025).
\newblock \urlprefix\url{https://arxiv.org/abs/2507.09089}

\bibitem{digitalpaktschule2026bilanz}
{BMBFSFJ}, {BMK}: Bilanzbericht digitalpakt schule 2019 -- 2025.
\newblock Tech. rep., Bundesministerium für Bildung, Familie, Senioren, Frauen
  und Jugend (2026).
\newblock \urlprefix\url{https://digitalpaktschule.de}.
\newblock Abgerufen am 3. September 2026

\bibitem{bria2025eurostack}
Bria, F., Timmers, P., Gernone, F.: {EuroStack} -- a european alternative for
  digital sovereignty.
\newblock Tech. rep., Bertelsmann Stiftung, G{\"u}tersloh (2025).
\newblock \doi{10.11586/2025006}

\bibitem{itea2026vendorlockin}
Bright, S., Emeny, W., Jaeger, A., Larson, A., Lueck, J.P., Soltys, M.:
  Avoiding vendor lock-in in {AI} procurement.
\newblock ITEA Journal \textbf{47}(2) (2026).
\newblock \doi{10.61278/itea.47.2.1002}

\bibitem{broeders2023search}
Broeders, D., Cristiano, F., Kaminska, M.: In search of digital sovereignty and
  strategic autonomy: Normative power {Europe} to the test of its geopolitical
  ambitions.
\newblock Journal of Common Market Studies \textbf{61}(5), 1261--1280 (2023).
\newblock \doi{10.1111/jcms.13462}

\bibitem{brynjolfsson2025generative}
Brynjolfsson, E., Li, D., Raymond, L.: Generative ai at work.
\newblock The Quarterly Journal of Economics \textbf{140}(2), 889--942 (2025)

\bibitem{brynjolfsson2021productivity}
Brynjolfsson, E., Rock, D., Syverson, C.: The productivity j-curve: How
  intangibles complement general purpose technologies.
\newblock American Economic Journal: Macroeconomics \textbf{13}(1), 333--372
  (2021).
\newblock \doi{10.1257/mac.20180386}

\bibitem{calvino2025generative}
Calvino, F., Haerle, D., Liu, S.: Is generative ai a general purpose
  technology?: Implications for productivity and policy.
\newblock OECD Artificial Intelligence Papers  (2025)

\bibitem{chester2023infrastructure}
Chester, M.V., Allenby, B.: Infrastructure and the cognitive ecosystem: an
  irrevocable transformation.
\newblock Environmental Research: Infrastructure and Sustainability
  \textbf{3}(3), 033002 (2023)

\bibitem{nato1969software}
Committee, N.S.: Software engineering: Report (1969)

\bibitem{cotroneo2025human}
Cotroneo, D., Improta, C., Liguori, P.: Human-written vs. ai-generated code: A
  large-scale study of defects, vulnerabilities, and complexity.
\newblock In: 2025 IEEE 36th International Symposium on Software Reliability
  Engineering (ISSRE), pp. 252--263. IEEE (2025)

\bibitem{cottier2025llm}
Cottier, B., Snodin, B., Owen, D., Adamczewski, T.: Llm inference prices have
  fallen rapidly but unequally across tasks, 2025.
\newblock Epoch AI, https://epoch.ai/data-insights/llm-inference-price-trends.
  Accessed Aug. 10, 2026  (2025)

\bibitem{eucouncil2025summit}
{Council of the European Union}: Summit on european digital sovereignty
  (berlin, 18 november 2025) -- information from france and germany.
\newblock Council Document 16186/25 (2025).
\newblock
  \urlprefix\url{https://data.consilium.europa.eu/doc/document/ST-16186-2025-INIT/en/pdf}.
\newblock Accessed on Sept 3, 2026

\bibitem{cui2026effects}
Cui, Z.K., Demirer, M., Jaffe, S., Musolff, L., Peng, S., Salz, T.: The effects
  of generative ai on high-skilled work: Evidence from three field experiments
  with software developers.
\newblock Management Science  (2026).
\newblock \doi{10.1287/mnsc.2025.00535}

\bibitem{dhanorkar2026human}
Dhanorkar, S., Passi, S., Vorvoreanu, M.: Human oversight of agentic systems in
  practice: Examining the oversight work, challenges, and heuristics of
  developers using software agents.
\newblock In: The 2026 ACM Conference on Fairness, Accountability, and
  Transparency, pp. 6438--6465 (2026)

\bibitem{drucker1994visible}
Drucker, J.: The visible word: experimental typography and modern art,
  1909-1923.
\newblock University of Chicago Press (1994)

\bibitem{eucom2026techsovereignty}
{European Commission}: Communication on european tech sovereignty, accompanied
  by an {EU} open source strategy.
\newblock COM(2026) 503 (2026).
\newblock
  \urlprefix\url{https://digital-strategy.ec.europa.eu/en/library/communication-european-tech-sovereignty-accompanied-eu-open-source-strategy}.
\newblock Accessed on Sept 3, 2026

\bibitem{falkner2024digital}
Falkner, G., Heidebrecht, S., Obendiek, A., Seidl, T.: Digital sovereignty --
  rhetoric and reality.
\newblock Journal of European Public Policy \textbf{31}(8), 2099--2120 (2024).
\newblock \doi{10.1080/13501763.2024.2358984}

\bibitem{floridi2020fight}
Floridi, L.: The fight for digital sovereignty: What it is, and why it matters,
  especially for the {EU}.
\newblock Philosophy \& Technology \textbf{33}(3), 369--378 (2020).
\newblock \doi{10.1007/s13347-020-00423-6}

\bibitem{fratini2024digital}
Fratini, S., Hine, E., Novelli, C., Roberts, H., Floridi, L.: Digital
  sovereignty: A descriptive analysis and a critical evaluation of existing
  models.
\newblock Digital Society \textbf{3}(59) (2024)

\bibitem{gi2026aikbse}
{Gesellschaft f{\"u}r Informatik e.V.}: Ki-basiertes software-engineering als
  schl{\"u}sseltechnologie digitaler souver{\"a}nit{\"a}t.
\newblock Policy Brief (2026).
\newblock
  \urlprefix\url{https://gi.de/meldung/mehr-als-vibe-coding-gi-veroeffentlicht-policy-brief-zu-ki-basiertem-software-engineering}

\bibitem{gundala2026SDD}
Gundala, M.: Spec-driven development as the governance backbone of the
  ai-driven development lifecycle (ai-dlc): A use case for enterprise-grade
  agentic software delivery.
\newblock International Journal of Scientific Research in Engineering and
  Management (IJSREM) \textbf{10}(7) (2026)

\bibitem{gundlach2025price}
Gundlach, H., Lynch, J., Mertens, M., Thompson, N.: The price of progress:
  Algorithmic efficiency and the falling cost of ai inference.
\newblock In: NeurIPS 2025 Workshop on Evaluating the Evolving LLM Lifecycle:
  Benchmarks, Emergent Abilities, and Scaling (2025)

\bibitem{gururaja2023build}
Gururaja, S., Bertsch, A., Na, C., Widder, D., Strubell, E.: To build our
  future, we must know our past: Contextualizing paradigm shifts in natural
  language processing.
\newblock In: Proceedings of the 2023 Conference on Empirical Methods in
  Natural Language Processing, pp. 13310--13325 (2023)

\bibitem{hassan2026agentic}
Hassan, A.E., Li, H., Lin, D., Adams, B., Chen, T.H., Kashiwa, Y., Qiu, D.:
  Agentic software engineering: Foundational pillars and a research roadmap
  (2026).
\newblock \urlprefix\url{https://arxiv.org/abs/2509.06216}

\bibitem{hooker2021hardware}
Hooker, S.: The hardware lottery.
\newblock Communications of the ACM \textbf{64}(12), 58--65 (2021).
\newblock \doi{10.1145/3467017}

\bibitem{IEEE610121990}
{IEEE Computer Society}: IEEE Standard Glossary of Software Engineering
  Terminology.
\newblock IEEE (1990).
\newblock \doi{10.1109/IEEESTD.1990.101064}

\bibitem{jevons1934william}
Jevons, H.W., Jevons, H.S.: William stanley jevons.
\newblock Econometrica, Journal of the Econometric Society pp. 225--237 (1934)

\bibitem{kaur2017interoperability}
Kaur, K., Sharma, D.S., Kahlon, D.K.S.: Interoperability and portability
  approaches in inter-connected clouds: A review.
\newblock ACM Computing Surveys (CSUR) \textbf{50}(4), 1--40 (2017)

\bibitem{korinek2025concentrating}
Korinek, A., Vipra, J.: Concentrating intelligence: scaling and market
  structure in artificial intelligence.
\newblock Economic Policy \textbf{40}(121), 225--256 (2025)

\bibitem{krizhevsky2012imagenet}
Krizhevsky, A., Sutskever, I., Hinton, G.E.: Imagenet classification with deep
  convolutional neural networks.
\newblock In: Advances in Neural Information Processing Systems, vol.~25 (2012)

\bibitem{lambach2022narratives}
Lambach, D., Oppermann, K.: Narratives of digital sovereignty in {German}
  political discourse.
\newblock Governance \textbf{36}(3), 693--709 (2022).
\newblock \doi{10.1111/gove.12690}

\bibitem{li2019dynamic}
Li, T.C., Chan, Y.E.: Dynamic information technology capability: Concept
  definition and framework development.
\newblock The Journal of Strategic Information Systems \textbf{28}(4), 101575
  (2019).
\newblock \doi{https://doi.org/10.1016/j.jsis.2019.101575}.
\newblock
  \urlprefix\url{https://www.sciencedirect.com/science/article/pii/S0963868717301415}

\bibitem{liu2026debt}
Liu, Y., Widyasari, R., Zhao, Y., Irsan, I.C., Chen, J., Lo, D.: Debt behind
  the ai boom: A large-scale empirical study of ai-generated code in the wild
  (2026).
\newblock \urlprefix\url{https://arxiv.org/abs/2603.28592}

\bibitem{mandl2026ai}
Mandl, P., Mandl, P.: Ai-driven software development: A pragmatic path to
  agentic development processes (2026).
\newblock \urlprefix\url{https://arxiv.org/abs/2606.15283}

\bibitem{moore1965cramming}
Moore, G.E., et~al.: Cramming more components onto integrated circuits (1965)

\bibitem{moore1975progress}
Moore, G.E., et~al.: Progress in digital integrated electronics.
\newblock In: Electron devices meeting, vol.~21, pp. 11--13. Washington, DC
  (1975)

\bibitem{nordhaus2007two}
Nordhaus, W.D.: Two centuries of productivity growth in computing.
\newblock The Journal of Economic History \textbf{67}(1), 128--159 (2007)

\bibitem{opara2016critical}
Opara-Martins, J., Sahandi, R., Tian, F.: Critical analysis of vendor lock-in
  and its impact on cloud computing migration: a business perspective.
\newblock Journal of Cloud Computing \textbf{5}(1), 4 (2016)

\bibitem{perry2023users}
Perry, N., Srivastava, M., Kumar, D., Boneh, D.: Do users write more insecure
  code with ai assistants?
\newblock In: Proceedings of the 2023 ACM SIGSAC conference on computer and
  communications security, pp. 2785--2799 (2023)

\bibitem{salim2026tokenomics}
Salim, M., Latendresse, J., Khatoonabadi, H., Shihab, E.: Tokenomics:
  Quantifying where tokens are used in agentic software engineering (2026).
\newblock \urlprefix\url{https://arxiv.org/abs/2601.14470}

\bibitem{santaniello2025attributes}
Santaniello, M.: Attributes of digital sovereignty: A conceptual framework.
\newblock Geopolitics \textbf{31}(2), 788--809 (2026)

\bibitem{Schieferdecker2025}
Schieferdecker, I.K.: The power of models for software engineering.
\newblock In: M.~Hinchey, B.~Steffen (eds.) The Combined Power of Research,
  Education, and Dissemination: Essays Dedicated to Tiziana Margaria on the
  Occasion of Her 60th Birthday, pp. 67--80. Springer Nature Switzerland, Cham
  (2025).
\newblock \doi{10.1007/978-3-031-73887-6_7}.
\newblock \urlprefix\url{https://doi.org/10.1007/978-3-031-73887-6_7}

\bibitem{Schieferdecker2026_AI4SE}
Schieferdecker, I.K.: Augmenting software engineering with ai. the ai4se
  taxonomy and its use.
\newblock Innovations in Systems and Software Engineering \textbf{22}(2), 29
  (2026)

\bibitem{sorrell2009jevons}
Sorrell, S.: Jevons’ paradox revisited: The evidence for backfire from
  improved energy efficiency.
\newblock Energy policy \textbf{37}(4), 1456--1469 (2009)

\bibitem{teece2007explicating}
Teece, D.J.: Explicating dynamic capabilities: the nature and microfoundations
  of (sustainable) enterprise performance.
\newblock Strategic management journal \textbf{28}(13), 1319--1350 (2007)

\bibitem{vella2026impact}
Vella, A., Blincoe, K.: The impact of ai coding assistants on software
  engineering: A longitudinal study (2026).
\newblock \urlprefix\url{https://arxiv.org/abs/2605.23135}

\bibitem{verhoef2021digital}
Verhoef, P.C., Broekhuizen, T., Bart, Y., Bhattacharya, A., {Qi Dong}, J.,
  Fabian, N., Haenlein, M.: Digital transformation: A multidisciplinary
  reflection and research agenda.
\newblock Journal of Business Research \textbf{122}, 889--901 (2021).
\newblock \doi{https://doi.org/10.1016/j.jbusres.2019.09.022}.
\newblock
  \urlprefix\url{https://www.sciencedirect.com/science/article/pii/S0148296319305478}

\bibitem{SWEBOK2024}
Washizaki, H. (ed.): Guide to the Software Engineering Body of Knowledge
  (SWEBOK Guide), Version 4.0.
\newblock IEEE Computer Society (2024)

\bibitem{SRIA2026_Matrix}
WG, S.R.: Securing the digital foundation: A strategic research innovation
  agenda (sria).
\newblock Tech. rep., Informatics Europe and ERCIM (2026)

\bibitem{yang2024sweagent}
Yang, J., Jimenez, C.E., Wettig, A., Lieret, K., Yao, S., Narasimhan, K.,
  Press, O.: Swe-agent: Agent-computer interfaces enable automated software
  engineering (2024).
\newblock \urlprefix\url{https://arxiv.org/abs/2405.15793}

\bibitem{zhang2026economics}
Zhang, Y., Zhang, T.: The economics of digital intelligence capital: Endogenous
  depreciation and the structural jevons paradox (2026).
\newblock \urlprefix\url{https://arxiv.org/abs/2601.12339}

\bibitem{zhu2026designing}
Zhu, L., Lu, Q., Ding, M., Lee, S.U., Wang, C.: Designing meaningful human
  oversight in ai.
\newblock AI and Ethics \textbf{6}(3), 286 (2026)

\end{thebibliography}
\endgroup}
\makeatother

\begin{document}

\selectlanguage{english}

\title*{When Digitalization Transforms Itself: \newline AI, Software, and the Next Technical Order}
\titlerunning{Digital Self-Transformation}
\author{Ina K. Schieferdecker}
\institute{Technische Universität Berlin \at Einsteinufer 25, 10587 Berlin, Germany, \email{ina.schieferdecker@tu-berlin.de}}

\maketitle

\abstract{Agentic artificial intelligence (AI) marks a new phase of digitalization: digitalization is beginning to act back upon its own technical production base. Whereas earlier phases aimed at digitizing analog information, automating processes, and building digital value networks, AI is increasingly taking over tasks in the development of digital systems themselves. In software engineering (SE) in particular, agentic AI systems can already plan, execute, check, and iteratively refine development tasks across multiple steps.
Digitalization thus becomes recursive: digital systems no longer merely support the digitalization of other sectors, but are increasingly developed and evolved by digital systems themselves. The technical production base of digitalization thereby becomes the object of its own transformation. This changes not only the division of labor between humans and machines in SE, but also the role of SE as a whole. This chapter analyzes this socio-technical change and argues that SE is evolving from a primarily supporting engineering discipline into a central infrastructure of societal value creation as well as of technological and digital sovereignty.}

\medskip
\noindent\textit{Note: This English text is a translation of the German original, which follows as the second part of this document.}

\section{Introduction}
\label{sec:intro}

When artificial intelligence (AI) no longer merely digitalizes industries but begins to digitalize and act back upon its own technical production base, digitalization enters a new, recursive stage of development: \textbf{digital self-\hspace{0pt}transformation}. Whereas earlier phases aimed at digitizing analog information and automating processes, agentic AI is increasingly taking over tasks in the creation, verification, and evolution of digital systems themselves. Software thereby turns from an instrument of digitalization into its product -- and software engineering (SE) turns from a supporting craft into a strategic infrastructure of societal value creation and technological sovereignty.

This shifts the object of digitalization. It is no longer confined to information, processes, and value-creation structures, but extends to the processes through which digital systems themselves are created, verified, and changed. This \textbf{digital self-\hspace{0pt}transformation} goes beyond yet another stage of automating software development. It changes the production logic of software, the division of labor between humans and machines, and the requirements for quality assurance, accountability, and governance. At the same time, new economic and technological path dependencies emerge: the more software becomes the foundation of digital value creation, and the more its production is shaped by AI-based systems, the more strategic becomes the control over the technical, infrastructural, and knowledge-based conditions of its production. The ability to develop, verify, and evolve software thus itself becomes a question of technological agency.

At the center of this development, therefore, is not only the performance of generative or agentic AI, but the question of how software engineering (SE) must be redesigned under these conditions. As machines increasingly take over implementation and other development tasks, engineering value creation shifts from the manual production of code towards the precise articulation of system intentions, the architecture of complex systems, the definition and enforcement of quality goals, as well as verification and validation (V\&V) and governance. SE thus becomes the object of digitalization in two respects: on the one hand, its processes and activities are themselves increasingly transformed by AI systems; on the other hand, SE becomes the central capability for designing, controlling, and evolving the emerging digital production systems. SE thereby gains a strategic importance that goes beyond its traditional role as a supporting engineering discipline.

From this perspective, the chapter examines the \textbf{fourth phase of digitalization} as a process of digital self-\hspace{0pt}transformation. It first considers the technical and economic foundations of the emerging mode of software production in Section~\ref{sec:phases}, then analyzes the changes in software production and SE in Section~\ref{sec:software}, and subsequently examines the change in the cognitive marginal costs of software production in Section~\ref{sec:marginalcosts}, the socio-technical dynamics of software production in Section~\ref{sec:dynamics}, and economic path dependencies as well as possible design options in Section~\ref{sec:design}. The central question is what role agentic SE plays for future digital value creation and what significance it has for the technological and digital sovereignty of Germany and Europe. A summary and outlook in Section~\ref{sec:summary} conclude the chapter.

\section{The Phases of Digitalization}
\label{sec:phases}

Digitalization has continuously expanded its scope over several phases (see Table~\ref{tab:phases}\footnote{For an explanation of the last column of this table, marginal costs, see Section~\ref{sec:marginalcosts}.}). The digitization of analog information was followed by the digitalization and automation of processes, and then by the digital networking and transformation of value-creation structures. With the advent of generative and, in particular, agentic artificial intelligence (AI), digitalization increasingly digitalizes itself in a recursive manner.

The first three phases of digitalization form the technological and economic development path against whose background this new quality becomes visible. In the first phase, technical digitization, analog information was converted into digital representations. With the beginning of the computer age, for example, paper files, accounting ledgers, or other analog records were transferred into data structures that could be processed by software. The focus was on the digital representation of existing structures.

In the second phase, process digitalization, the scope was extended to work and business processes. Software and digital networking enabled their automation and optimization, for example through e-mail, ERP systems, or online shops. While this could fundamentally change how processes were carried out, the underlying business models often remained largely intact.

The third phase, digital transformation, finally encompassed not just individual processes but increasingly entire value-creation structures. Platforms, cloud computing, and the Internet of Things enabled new forms of networking, coordination, and scaling. This gave rise to new business models and organizational forms. Digitalization thus itself became a central driver of economic and societal change.

The characteristic feature of the fourth phase is the recursiveness of digitalization, i.e., digital self-\hspace{0pt}transformation: the technical systems and processes through which digitalization is realized become themselves the object of digital transformation. Agentic AI in particular enables development and optimization processes in which digital systems interpret goals, plan subtasks, use tools, evaluate results, and carry out further actions based on feedback. The qualitative break of the fourth phase lies in its inherent recursiveness: instead of `merely' permeating external industries, digitalization digitalizes its own genesis. The focus shifts to those machine and human processes that create and change software. Software thus transcends its role as a mere tool of digital transformation and becomes an independent, agentically generated result of a self-transforming production base.

\begin{table}
\centering
\renewcommand{\arraystretch}{1.2}
\setlength{\tabcolsep}{2pt}
\begin{tabular}{|>{\raggedright\arraybackslash}p{0.8cm}|>{\raggedright\arraybackslash}p{1.8cm}|>{\raggedright\arraybackslash}p{1.6cm}|>{\raggedright\arraybackslash}p{2.1cm}|>{\raggedright\arraybackslash}p{2.5cm}|>{\raggedright\arraybackslash}p{2cm}|}
\hline
    \textbf{Phase} &
    \textbf{Name} &
    \textbf{When} &
    \textbf{Object} &
    \textbf{Characteristics} &
    \textbf{Declining marginal costs} \\
\hline
    1 &
    Technical digitization &
    1960s--1990s &
    Information &
    Analog information is represented and processed digitally. &
    Storage and processing \\
\hline
    2 &
    Process digitalization &
    1990s--2010s &
    Processes &
    Digital processes are automated and optimized. &
    Process execution \\
\hline
    3 &
    Digital\newline transformation &
    2010s--2020s &
    Value-creation structures &
    Networked digital processes enable new business models and organizational forms. &
    Communication and coordination \\
\hline
    4 &
    Digital\newline self-trans\-for\-ma\-tion &
    since the 2020s &
    Technical\newline production base of digitalization &
    Digital systems increasingly take over the development, verification, adaptation, and optimization of digital systems themselves. &
    Cognitive synthesis \\
\hline
\end{tabular}
\caption{The phases of digitalization}
\label{tab:phases}
\end{table}

The fourth phase can be observed particularly clearly in the relationship between products, processes, and their production. In the media industry, for example, digitalization first led to digital storage media, then to the digitalization of distribution and business processes, and finally to platform-based and data-driven business models. Generative AI now adds a further quality: content can increasingly be generated situationally and individually. The boundary between production and reception shifts, because digital content is no longer exclusively produced in advance and consumed afterwards, but in part only comes into being in the process of its use.

A comparable change can be seen in industrial automation. Whereas early automation relied on deterministically programmed sequences, networked systems first enabled the integration of production and business processes. Digital twins, the Internet of Things, and predictive maintenance extended these capabilities with data-based analysis and optimization. The optimization goals and decision rules, however, basically remained defined by humans. In the fourth phase this boundary shifts: AI-based systems can increasingly interpret production states, generate options for action, evaluate their outcomes, and determine their next actions on that basis. What matters here is not the full autonomy of individual machines, but the emergence of recursive feedback loops in which digital systems participate in changing their own processes.

\subsection{Digital Self-Transformation}
\label{sec:selftransformation}

The characteristic feature of the fourth phase of digitalization is that digitalization acts back upon its own technical production base (see Table~\ref{tab:selftransformation}). Agentic AI is a key technological enabler of this development. In SE it manifests itself as agentic SE, in which AI agents increasingly take over multi-step development tasks across the software life cycle and coordinate their actions on the basis of feedback. The decisive change, therefore, is not only that AI generates code, but that it is increasingly involved in the processes through which software is planned, developed, verified, and evolved.

\begin{table}
\centering
\renewcommand{\arraystretch}{1.2}
\setlength{\tabcolsep}{2pt}
\begin{tabular}{|p{3.3cm}|>{\raggedright\arraybackslash}p{8.5cm}|}
\hline
    \textbf{Feature} &
    \textbf{Digital self-\hspace{0pt}transformation} \\
\hline
    Object &
    Development, adaptation, and optimization of digital systems themselves\\
\hline
    Technological basis &
    AI models, agentic AI, AI-augmented SE assets, as well as training, simulation, and evaluation environments\\
\hline
    Mode of production
    & Recursive, feedback-based, and increasingly agentically executed development processes\\
\hline
    Products and services
    & Adaptive, individualized, self-optimizing, and increasingly autonomous systems\\
\hline
    Organization &
    Human--AI and multi-agent collaboration\\
\hline
    Value creation &
    Cognitive industrialization and high degrees of individualization\\
\hline
    Governance &
    Continuous V\&V, monitoring, and control\\
\hline
Metrics & Quality, robustness, security, sustainability, trustworthiness, as well as cost and productivity\\
\hline
\end{tabular}
\caption{Characteristics of the fourth phase of digitalization: digital self-\hspace{0pt}transformation}
\label{tab:selftransformation}
\end{table}

This analysis builds on the distinction between digitization, digitalization, and digital transformation made by Verhoef et al.~\cite{verhoef2021digital} and extends this perspective by a fourth, currently emerging phase. While Verhoef et al.~\cite{verhoef2021digital} consider the three phases primarily from the perspective of the transformation of companies, resources, organizational structures, and business models, in the fourth phase the object of digitalization itself shifts.

In the following, the term \textbf{SE asset} is used, first, to point to the difference between classical SE artifacts and agentic assets in SE. Classical SE artifacts such as requirements documents, architecture diagrams, code, or test cases are typically persistent, versioned results of a work step in SE that are authorized by humans~\cite{SWEBOK2024}. In agentic SE, by contrast, intermediate products such as agent plans, context and memory objects, or tentatively generated and discarded code versions continuously emerge; they are machine-generated, iteratively modified or overwritten, and partly ephemeral~\cite{hassan2026agentic}. The term is furthermore used to denote both AI-generated~\cite{mandl2026ai} and AI-used intermediate and final products. It thus allows a compact description of essential dynamics of SE in digital self-\hspace{0pt}transformation. The term SE asset was chosen to distinguish it from AI-augmented SE in~\cite{Schieferdecker2026_AI4SE} and from AI-generated artifacts in~\cite{mandl2026ai}.

Even though the characteristics of the fourth phase of digitalization can only be fully determined as it develops further, essential features are already emerging today. A central change concerns the mode of production of software: foundation models (AI models for short), SE assets, as well as training, simulation, and evaluation environments form a new resource and production base. In SE, this base condenses into a constellation of big data, big code, and big (software) models: specifications, models, and code are no longer merely results of software development, but themselves become machine-interpretable resources for its analysis, generation, and evolution~\cite{Schieferdecker2026_AI4SE}.

For example, with the currently emerging so-called spec-driven development, specifications increasingly become control and reference points for AI agents, which derive plans, implementations, and tests from them and can verify results against specified properties~\cite{gundala2026SDD}. In the concept of so-called loop engineering, this approach is extended to continuous agentic feedback loops in which AI agents plan, act, observe and verify results, and determine their next actions on that basis~\cite{hassan2026agentic}. AI agents can thus increasingly take over tasks across several phases of the software life cycle and, in doing so, generate new SE assets themselves. They are increasingly stepping out of their role as consumers and optimizers of SE assets and are becoming their producers.

The resulting organizational structures are socio-technical in nature. Humans increasingly define goals, constraints, quality requirements, and responsibilities, while AI agents at least partially take over planning, implementation, testing, V\&V, and optimization. The central engineering task thus shifts from the manual creation of individual SE assets towards the design and control of the processes in which humans and machines jointly create and evolve software.

The economic logic changes as well. Value creation increasingly targets highly individualized, adaptive, and continuously optimized products and services. At the same time, the focus of optimization shifts from cost and productivity to a multidimensional interplay of cost, quality, security, sustainability, and trustworthiness. The more systems can influence their own development and optimization processes, the more important it becomes to determine what is to be optimized. Defining goals, constraints, and acceptable trade-offs becomes a central technical, organizational, and societal design task.

Digital self-\hspace{0pt}transformation thus does not denote a completed phase of digitalization, but a currently emerging constellation whose reach and limits are still open. Its decisive feature is recursiveness: digitalization is not only advanced further, but increasingly changes the conditions under which digitalization itself is produced.

\section{Software Production in Transition}
\label{sec:software}

Software is the central technical carrier of digitalization. Almost all digital products, services, and processes are based on software. Its production, however, is far more than writing code. Already with the first complex software systems it became evident that requirements must be captured systematically, solution structures designed, implementations created, quality checked, and changes controlled throughout the life cycle. With the growing complexity and criticality of software-based systems, the ability to systematically integrate these different activities therefore became decisive. Software production evolved from a primarily individual programming activity into a complex engineering process.

\begin{figure}
    \centering
    \includegraphics[width=\linewidth]{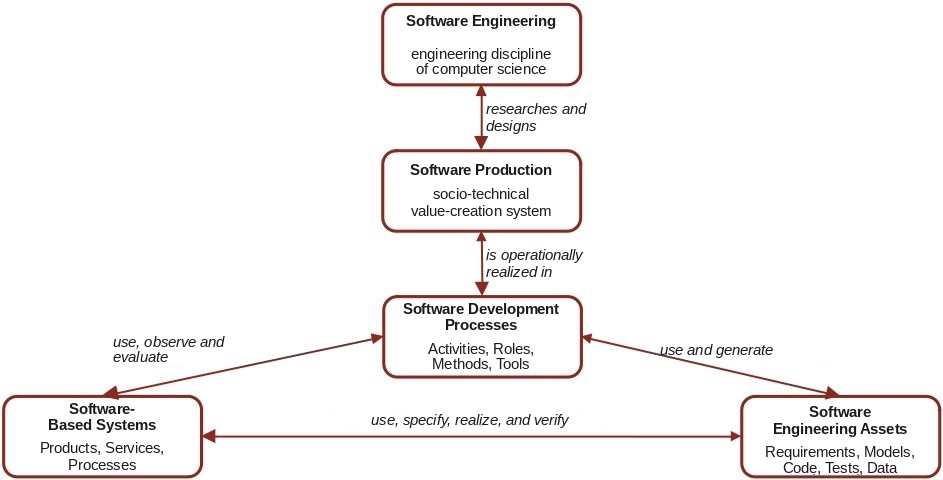}
    \caption{Relationships between SE, software production, software development processes, SE assets, and software-based systems}
    \label{fig:softwareengineering}
\end{figure}

In the context of the software crisis of the late 1960s~\cite{nato1969software}, this development led to the emergence of SE as an independent discipline of computer science~\cite{SWEBOK2024}. SE denotes the systematic, disciplined, and quantifiable application of engineering principles to the specification, development, evolution, and operation of software-based systems~\cite{IEEE610121990}. The goal is to make software not only functional, but also reliable, secure, maintainable, traceable, and economically developable.

To this end, SE organizes the entire development process: from the definition of goals and requirements through architecture and implementation to testing, V\&V, operation, and evolution. In the process, various SE assets are created and used. These include, for example, requirements, specifications and models, architectures, code, tests, data, as well as logs, traces, and documentation. They are the structured knowledge and working foundations on which the development, quality assurance, and evolution of software-based systems are built (see Figure~\ref{fig:softwareengineering}).

With the growing importance of software-based systems, it is thus not only software itself that becomes a central societal and economic infrastructure. The ability to systematically specify, develop, verify, and evolve software also becomes a strategic capability. SE designs and masters this software production and develops the methods, processes, and tools required for it.

This is precisely where the fourth phase of digitalization comes in (see Table~\ref{tab:assets}): whereas digital transformation has so far digitalized other industries, the software production industry itself is now increasingly becoming the object of digitalization.

This change concerns more than the automated generation of program code. In classical SE, SE assets are predominantly created through human engineering work. Requirements are formulated, architectures designed, implementations created, and tests developed. Automation methods such as model-driven SE~\cite{Schieferdecker2025} support individual activities, but the fundamental work in the development process remains predominantly determined by humans.

In agentic SE, the division of labor changes. AI agents can connect several engineering activities: they can analyze requirements, generate plans, invoke development tools, generate and modify code, run tests, evaluate results, and initiate further development steps on that basis. SE assets are thus no longer merely results of software development, but can themselves become dynamic resources that are generated, interpreted, modified, and recursively used by AI agents for further development steps.

\begin{table}
    \centering
    \renewcommand{\arraystretch}{1.2}
    \setlength{\tabcolsep}{2pt}
    \begin{tabular}{|>{\raggedright\arraybackslash}p{2cm}|
                    >{\raggedright\arraybackslash}p{3cm}|
                    >{\raggedright\arraybackslash}p{3cm}|
                    >{\raggedright\arraybackslash}p{3cm}|}
        \hline
        \textbf{SE asset} & \textbf{Classical SE} & \textbf{Agentic SE} & \textbf{Agentic loop}\\
        \hline
        Requirements &
        Goal definition &
        Agent steering &
        Goal adaptation\\
        \hline
        Specifications / models &
        System description &
        Agent context and steering &
        Model generation and evolution\\
        \hline
        Project plans / tasks &
        Structuring of work &
        Agent planning and orchestration &
        Dynamic task adaptation\\
        \hline
        Code &
        Implementation &
        Generation and analysis &
        Continuous evolution\\
        \hline
        Tests &
        V\&V &
        Feedback and evidence &
        Automatic test generation and optimization\\
        \hline
        Data &
        Information and knowledge base &
        Context and evidence &
        Continuous enrichment\\
        \hline
        Logs / traces &
        Diagnosis and traceability &
        Experience and evaluation data &
        Basis for learning and optimization\\
        \hline
        Policies / engineering guidelines &
        Guidelines &
        Agent directives &
        Declarative process control\\
        \hline
    \end{tabular}
    \caption{SE assets in the transition from classical to agentic SE -- from static to dynamic artifacts}
    \label{tab:assets}
\end{table}

The change can be illustrated by the different roles of SE assets. In classical SE, requirements and the resulting specifications and models primarily serve to describe the goals and properties of a system for humans and development tools. In agentic SE, they increasingly become machine-interpretable control and reference points: AI agents can derive project plans from them, generate implementations, and check results against given properties. In spec-driven development (see Section~\ref{sec:phases}), specifications take on a new meaning in agentic SE processes: they not only describe what a system is supposed to do, but at the same time serve as persistently available context and as a reference for evaluating the SE assets generated by agentic AI systems.

Code likewise changes its role. It is no longer exclusively the result of human implementation, but can be generated, analyzed, tested, and modified by AI agents. Code thus increasingly becomes an intermediate product within an iterative engineering process. Tests and evaluations are thereby transformed from downstream quality-assurance activities into continuous sources of evidence and feedback, on the basis of which AI agents make subsequent development decisions and iteratively refine their results.

In so-called loop engineering, this feedback becomes a central organizing principle of agentic software production. AI agents plan actions, carry them out, observe their outcomes, evaluate these outcomes, and determine further tasks on that basis~\cite{hassan2026agentic}. Development thus becomes an iterative process in which the results of one development activity become the starting point of further development activities. AI agents can increasingly create and modify SE assets themselves; as described in Section~\ref{sec:selftransformation}, they thereby increasingly become producers of SE assets.

Data as well as logs and traces acquire additional significance in this process. They no longer merely document past system states or support fault diagnosis, but can feed into subsequent project decisions as context, experience, and evidence resources. Policies and engineering guidelines, in turn, will serve as machine-interpretable directives for the execution and limitation of agentic engineering processes. They thus define not only constraints on the emerging system, but increasingly also constraints on its development.

Agentic SE therefore does not merely change individual SE tools. It changes the relationship between the activities of the development process and the SE assets created in it. What in classical SE is predominantly the result of a development activity can, in agentic SE, simultaneously be the starting point and control basis for further development activities. SE assets thus increasingly form the foundation of recursive software development processes.

These developments in software production explain the particular significance of agentic SE for the fourth phase of digitalization. The decisive change is not that machines generate software for the first time. Automated code generation, model-based development, and other forms of SE automation have existed for decades. What is new is rather the increasing integration of different SE activities in agentic, feedback-based processes. As a result, software can increasingly not only be generated automatically, but also be analyzed, checked, modified, and evolved within a coherent process.

\subsection{From Manual Code Production to the Design of Complex Software-Based Socio-Technical Systems}

With the increasing automation of code generation, the focus of engineering value creation in SE therefore shifts. It is less about manually writing individual lines of code and increasingly about the precise articulation of system intentions, the architecture of complex software-based systems, the definition of quality goals, V\&V, and the design and governance of human--AI and multi-agent production processes.

An instructive historical parallel is offered by the development of the typographer or compositor~\cite{drucker1994visible}. Classical compositors mastered a specialized craft: using lead type and later phototypesetting, they turned texts into a printable typographic form. With the introduction of desktop publishing in the 1980s, essential parts of this production work were digitalized and, at the same time, made accessible to other actors. Typographic design could now be done directly at the computer; the technical separation between authors, designers, typesetters, and prepress was partly dissolved. What disappeared was not the need for typography or design, but a considerable part of the compositor's specialized manual production work.
This historical development can be understood less as the disappearance of an entire profession than as a shift in its position in value creation. Machines and software tools took over standardizable production steps; human work shifted towards design, selection, composition, and quality judgment. Digitalization thus does not necessarily eliminate a task as a whole, but can change the previous separation between specialized production activities and overarching design.

A structurally similar development is emerging for SE. Generative AI, in particular large language models (LLMs), can already generate code from requirements formulated in natural language; at the same time, empirical studies show that the quality of such output does not automatically meet the requirements of professionally developed software (see also Section~\ref{sec:dynamics}). This points to a possible shift from code-centric to intent-centric SE: code turns from the primary development product into one element of a recursive engineering process.

The typographer analogy also makes clear why the question of whether AI might replace programmers falls short. What matters instead is which SE activities are automated, which new activities emerge, and which competencies are lifted to a higher level of abstraction. Just as desktop publishing largely automated typesetting without eliminating the need for typography, design, and visual communication, generative and agentic AI can largely automate the creation of code without eliminating the need for SE.

The difference, however, is fundamental: software is not merely a form of representing or processing information, but constitutes the technical foundation of executable and increasingly socially critical socio-technical systems. Faulty or insecure software can cause economic damage, endanger people, or impair critical infrastructures. With the increasing automation of code generation, responsibility for software quality therefore does not disappear. Rather, it increasingly shifts towards determining goals and quality requirements, monitoring and evaluating development results, and governing partially automated development processes.

This is precisely where the strategic significance of SE in the fourth phase of digitalization lies. When generative and agentic SE tools increasingly generate code and other SE assets, SE does not become less relevant. Rather, it becomes the discipline that orients, structures, limits, checks, and makes accountable this partially automated software production. Engineering value creation thus shifts from the direct production of code towards the design of intention, architecture, quality, and evolution of complex software-based socio-technical systems.

This shift, however, does not concern only the methods and tools of SE. It is also based on a change in the resources and conditions under which software is produced.

\subsection{Cognitive Infrastructures as a Factor of Production}
\label{subsec:infrastructure}

For decades, cognitive work in digitalization was tied to persons, teams, and organizations. Additional capacity therefore primarily required education, recruitment, collaboration, and knowledge building.

Generative and agentic AI change these conditions of production. Certain cognitive services can increasingly be provided, retrieved, and scaled digitally and integrated directly into work processes. Similar to how cloud computing decoupled computing power from local hardware~\cite{armbrust2009above}, AI enables the provision of certain cognitive services independently of the immediate availability of individual persons. In certain fields of application, cognitive capacity thus becomes an elastically deployable resource.

This technical elasticity forms the basis for a new form of digital infrastructure. \textbf{Cognitive infrastructure} here denotes technical, organizational, and institutional resources through which cognitive services such as knowledge processing, generation, planning, and problem solving can be provided as reusable and scalable services and integrated into digital work processes. The concept builds on the notion of cognitive infrastructure as an interplay of technologies, services, institutions, and products that provide functional elements of cognition~\cite{chester2023infrastructure}.

As a factor of production, such an infrastructure makes it possible to integrate cognitive capacity into production and decision-making processes at different scales, independently of the immediate availability of individual persons or teams. Its productive effect depends not only on the performance of the underlying AI, but equally on its availability, scalability, reliability, and embedding in organizational processes. Cognitive infrastructure thus creates the precondition for certain cognitive production services to be provided at substantially changed marginal costs. The resulting economic consequences for software production are considered in the following section.

\section{The Economics of Cognitive Marginal Costs}
\label{sec:marginalcosts}

The phases of digitalization can be distinguished not only by what is digitalized and with what, but also by which costs of digital services decline as a result of technological progress (see Table~\ref{tab:phases}). First, the costs of storing and processing information fell, then the costs of automated execution and networking of digital processes. With digital transformation, communication, coordination, and data-based prediction in particular became cheaper.

The fourth phase of digitalization targets a further stage of production: agentic AI increasingly lowers the marginal costs of certain cognitive services that have so far been provided predominantly by human labor. This also changes the cost structure of software production. In addition to the traditional expenditures for personnel, tools, and infrastructure, there are ongoing costs for AI model inference and agentic interactions. These arise not only in code generation, but also in planning, context processing, tool use, review, testing, and correction.

The significance of these variable costs is particularly evident in agentic SE: a recent empirical study of multi-agent systems in the software development life cycle (SDLC), for example, reports for the environment studied that on average 59.4\% of total token consumption was attributable to the iterative code review phase; 53.9\% of consumption was due to input tokens~\cite{salim2026tokenomics}. The costs of agentic software production thus arise not only in the generation of code, but throughout the entire iterative process of context processing, interaction, and evaluation.

This shifts the economic perspective on software production. What matters is not only how much the productivity of software development increases through AI, but how the declining marginal costs of cognitive services affect running costs, demand, and the required production resources.

\subsection{Lowering Cognitive Marginal Costs}
\label{subsec:lowering}

The first phase of digitalization initially reduced the costs of storing, copying, and processing digital information. Process digitalization subsequently lowered the costs of automated execution and reproduction of standardized digital processes. With digital transformation, networking, platforms, and data-based methods considerably reduced the costs of coordination and especially of data-based prediction. Agrawal et al. accordingly interpret the economic core of AI as a substantial reduction in the cost of prediction~\cite{agrawal2018prediction}.

The current phase of digital self-\hspace{0pt}transformation extends this development by a further dimension: agentic AI increasingly lowers the marginal costs of \textbf{cognitive synthesis}. Cognitive synthesis is understood here as the ability to generate new SE assets, recommendations, decisions, or action plans from goals, knowledge, and context, and to evaluate their suitability based on feedback. In the context of software production, this includes, for example, deriving architecture designs, code, tests, or project plans from requirements and existing system knowledge.

The reduction of digital marginal costs has a long technical history. As early as 1965, Moore described the long-term increase in the integration density of integrated circuits and revised his forecast in 1975~\cite{moore1965cramming,moore1975progress}. What matters here is less the specific doubling rate than the substantially declining cost of digital computing power over decades. Advances in semiconductor technology, computer architecture, software, and algorithms made digital processing continuously more powerful and less expensive~\cite{nordhaus2007two}. With generative and agentic AI, this development continues at a higher level of abstraction: not only computational operations but increasingly cognitive services become digitally executable and scalable.

\subsection{The Variable Costs of Agentic AI in Software Production}

For the following discussion (see Table~\ref{tab:costs}), two dimensions of software are distinguished: the mode of software creation and the mode of software execution. Software is referred to as \textbf{classically created} if its creation and evolution are carried out primarily through human software development, even if automated development tools or AI-based assistants are used. Software is referred to as \textbf{agentically created} if AI agents independently take over essential parts of its creation, verification, or evolution in multi-step work processes.
Independently of this, one must consider how the created software is executed. Classical software processes inputs on the basis of explicitly implemented program logic and thus follows an execution logic that is largely fixed at runtime. Agentic software, in contrast, uses AI models to plan and process tasks independently and may execute several inference, tool, and feedback steps to do so. ``Classical'' and ``agentic'' thus denote different characteristics of the mode of creation and of execution. Combining both dimensions yields four ideal-typical forms of software: classically created classical software (classical software for short), classically created agentic software, agentically created classical software, and agentically created agentic software (agentic software for short).

Declining marginal costs of cognitive services for software production do not mean that their provision becomes free of charge. Rather, the cost structure changes: while for classical software the major development expenditures are largely independent of the number of copies and duplication is possible at very low marginal costs, agentic software incurs variable costs for computing power, energy, and technical infrastructure with every additional inference (see Table~\ref{tab:costs}). This relationship is reinforced by the fact that in software (production), complex tasks are typically not handled by a single AI model call, but by sequences of inference steps, tool calls, and evaluation and feedback loops. The cost of a cognitive service thus depends not only on the underlying AI model, but also on the number and complexity of the processing steps required to provide it. For agentic software production, this means that its economics are increasingly determined not only by development costs but also by the running costs of agentic execution.

At the same time, the costs of this execution are falling considerably. The level of inference costs depends, among other things, on the complexity of the task, the number and type of AI model calls, and the hardware and software infrastructure used. Studies by Epoch AI show that inference prices for a given level of performance can fall by factors between about nine and several hundred within a year, depending on the task~\cite{cottier2025llm}. More recent analyses attribute this development to improvements in hardware and system efficiency as well as to algorithmic progress and increasing competition~\cite{gundlach2025price}. This gives rise to an economic dynamic characteristic of the fourth phase of digitalization: cognitive services remain associated with positive variable costs, but their marginal costs are falling rapidly. Cognitive processing can therefore be automated, repeated, and scaled to an ever greater extent. What matters, then, is not only the cost of an individual AI service, but the relationship between declining costs per unit of service and the resulting expansion of its economically meaningful use.

This development also changes classical software economics (see Table~\ref{tab:costs}). For classical software, the major development expenditures typically occur before use, while executing and duplicating an already developed program is possible at low marginal costs. For agentically created software, in contrast, production itself becomes a recurring, compute-intensive inference process: the creation, verification, and evolution of SE assets can be carried out by AI agents and thus incur variable costs for further production and change steps. Declining inference costs make this form of production increasingly economically attractive, but do not eliminate the associated follow-up costs.

Declining marginal costs of code generation can at the same time trigger an opposing cost effect: the more cheaply SE assets can be generated, the larger the amount of code that is generated and hence has to be reviewed, integrated, and maintained. A large-scale study of 304,362 verified AI-generated commits from 6,275 GitHub repositories identified 484,606 issues introduced by these commits; 89.1\% of them were code smells. In more than 15\% of the examined commits of the AI tools considered, at least one such issue was introduced. Of particular relevance is that 24.2\% of the tracked issues introduced in AI-generated commits still persisted in the latest revision examined~\cite{liu2026debt}. Declining costs of code generation therefore do not automatically lead to declining total costs of software production. Rather, they can shift the cost structure within the software life cycle: savings in code generation may be accompanied by additional expenditures for V\&V, integration, bug fixing, and long-term maintenance. The economic effect of agentic software production is thus determined not only by how cheaply code can be generated, but by how total costs develop over its life cycle.

\begin{table}
    \centering
    \renewcommand{\arraystretch}{1.2}
    \setlength{\tabcolsep}{2pt}
    \begin{tabular}{|>{\raggedright\arraybackslash}p{2.2cm}|>{\raggedright\arraybackslash}p{2.2cm}|>{\raggedright\arraybackslash}p{2.2cm}|>{\raggedright\arraybackslash}p{2.2cm}|>{\raggedright\arraybackslash}p{2.2cm}|}
    \hline
        \textbf{Phase} &
        \textbf{Classical software} &
        \textbf{Agentically created classical software} &
        \textbf{Classically created agentic software} &
        \textbf{Agentic software} \\
    \hline
        Development and evolution &
        predominantly fixed total costs &
        variable AI inference costs &
        predominantly fixed total costs &
        variable AI inference costs \\
    \hline
        Execution and use &
        low marginal costs &
        low marginal costs &
        variable AI inference costs &
        variable AI inference costs \\
    \hline
    \end{tabular}
    \caption{Simplified cost characteristics of different forms of software}
    \label{tab:costs}
\end{table}

\subsection{A Possible Expansion of Software Production}

Furthermore, declining marginal costs of software production do not necessarily lead to a reduction in the overall demand for software. Rather, productivity gains can trigger a rebound effect~\cite{jevons1934william} by making applications economically viable whose development would not have taken place at higher costs. In this sense, the Jevons paradox can be transferred to software production as a heuristic frame of reference: more efficient production can expand the use of a good instead of reducing it~\cite{sorrell2009jevons}.

For software, the expansion of the addressable application space is particularly relevant. Declining development costs not only make additional applications in existing markets economically viable, but also enable solutions for individual, situational, or temporary requirements whose development effort previously exceeded the expected benefit. The economically relevant boundary thus shifts from the question of whether a software solution is technically feasible to the question of for which additional purposes its development becomes economically worthwhile.

Agentic SE can reinforce this development by not only generating software more efficiently, but also supporting its adaptation to changing contexts. Software can thereby turn from an artifact developed in advance and subsequently largely static into an adaptable resource. First empirical studies already show substantial productivity gains from generative AI in certain knowledge-intensive activities~\cite{brynjolfsson2025generative}. Whether the resulting decline in production costs will lead in the long run to a disproportionate expansion of the overall demand for software, however, is an empirical question and is not settled by the productivity argument alone.

\subsection{The Shift of Economic Scarcities}

An expansion of software production at the same time changes the relative importance of the resources needed for software production and alters the structure of scarcities within software production. If the generation of software can increasingly be automated, pure production capacity loses importance as a differentiating factor~\cite{zhang2026economics}. At the same time, resources and capabilities required for selecting, evaluating, safeguarding, and embedding generated systems may become relatively scarcer.

These include, in particular, reliable data, computing and energy infrastructure, as well as competencies in V\&V, quality assurance, security, data protection, and governance. Added to this is human judgment in specifying goals, evaluating results, and deciding on the deployment of systems in concrete social and institutional contexts.

The economic development of software production is therefore not to be understood as a transition from scarcity to general abundance. Rather, relative scarcity shifts between different factors of production: while the costs of generating software fall and its supply can be expanded, complementary resources and capabilities may become relative bottlenecks. The increasing availability of code thus does not abolish economic scarcity, but partially relocates it within software production.

\section{Socio-Technical Dynamics of Software Production}
\label{sec:dynamics}

With agentic AI, the socio-technical organization of software production changes as well: the division of labor, organizational structures, and control mechanisms must be redesigned. This also raises the question of what the engineering achievement in software development will consist of in the future.

\begin{figure}
    \centering
    \includegraphics[width=\linewidth]{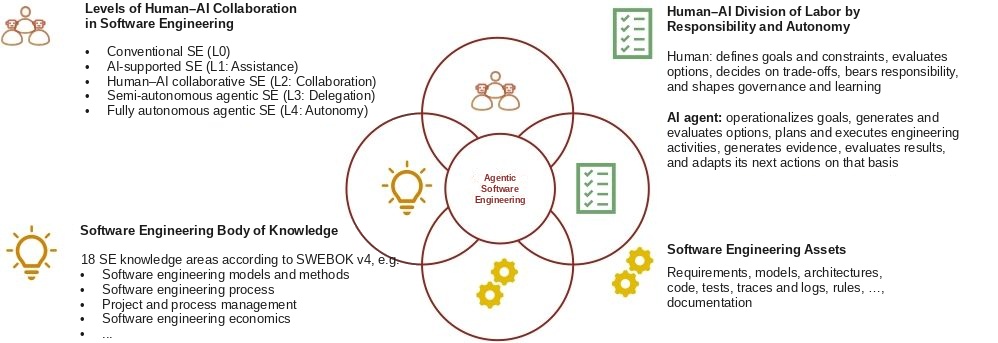}
    \caption{Agentic SE as a socio-technical system}
    \label{fig:agenticSE}
\end{figure}

Whereas earlier AI-based assistants in SE mainly supported individual activities such as the generation or transformation of SE artifacts, agentic SE systems can increasingly carry out coherent development tasks across multiple steps. They analyze requirements and code repositories, create plans, modify files, use software tools, run tests, and recursively adapt their approach based on the results. Systems such as SWE-agent demonstrate this shift from code assistance to the repository-wide handling of software engineering tasks~\cite{yang2024sweagent}.

This changes the division of labor between humans and machines: instead of performing individual activities, humans increasingly delegate coherent tasks to an agentic SE system and steer its scope of action through goals, context, constraints, and control mechanisms. Development thus shifts from occasional assistance through collaboration and delegation towards increasingly autonomous development activities (see Figure~\ref{fig:agenticSE}).

\subsection{From Programming to Supervisory Engineering}

Empirical studies on the use of AI coding assistants support the thesis of a shift towards increasingly autonomous development activities: a longitudinal study among professional software developers shows that 82\% of respondents spend less time writing code and that the activity as a whole is shifting from creation towards V\&V activities~\cite{vella2026impact}. The authors introduce the term \textbf{supervisory engineering work} for a form of engineering work that comprises the steering, evaluation, and correction of AI-generated SE assets.

Software engineers do not thereby become mere managers of AI agents, but increasingly take on responsibility for designing and monitoring the production process in which agentic AI systems execute (sub)tasks (semi-)autonomously. Supervisory engineering comprises in particular the definition of work assignments and constraints, the steering of agent workflows, the evaluation of their results, and the identification and correction of errors that are not immediately recognizable at the level of individual code components~\cite{vella2026impact}.

This also shifts the temporal and cognitive structure of development work. The direct production of code decreases, while review, V\&V, error analysis, and repeated interaction with AI agents gain in importance. Vella et al.~\cite{vella2026impact} reveal a remarkable productivity--experience paradox: 84\% of respondents reported productivity improvements both at the beginning and six months later, while at the same time the share of people who perceived a deterioration in at least one dimension of developer experience rose from 14\% to 27\%. Flow and perceived cognitive load in particular deteriorated, while feedback loops improved.

Furthermore, the results in~\cite{vella2026impact} put a purely output-oriented view of productivity into perspective. Higher speed in generating software does not necessarily mean a better work experience or greater effectiveness of the development process as a whole. In a different empirical setting, a randomized controlled trial even showed a 19\% longer completion time when the use of then-current AI coding tools was allowed~\cite{becker2025measuring}. At the same time, after completing the study, the developers estimated their own productivity gain at 20\% on average. The authors emphasize, however, that these results are to be understood as a context-dependent snapshot and cannot be generalized to software development as a whole~\cite{becker2025measuring}.

With the transformation of software production, the competence profile of software engineers changes as well. System understanding, precision in requirements, architectural competence, V\&V competence, and the ability to effectively orchestrate agentic AI systems in a given technical and organizational context become increasingly decisive.

The central change from classical SE to agentic SE thus consists less in a simple substitution of human development work than in a shift from programming to conception and supervision: engineers increasingly design the conditions under which agentic AI systems program software, and at the same time bear responsibility for the technical adequacy and quality of the result.

\subsection{The Reconfiguration of Teams and Organizational Structures}

When agentic AI systems take over a larger share of development activities, the organization of software production inevitably changes as well. Tasks that were previously distributed across several specialized roles can increasingly be coordinated by smaller human units that draw on a multitude of specialized AI agents. This can increase the scope of action of individual engineers.

However, no empirically validated new standard form of team organization can yet be derived from this development. The frequently voiced idea of highly integrated micro-teams or so-called super cells that handle large parts of the software stack with the help of (swarms of) AI agents is therefore to be understood, for now, as an organizational hypothesis. Its realization depends, among other things, on the reliability of the AI agents and their embedding in software production, the complexity of the software-based systems to be developed, the required domain expertise, and the regulatory and safety-related context.

A shift in the functions of SE roles and teams appears more likely, for the time being, than a general dissolution of their specializations. Platform and infrastructure teams could gain in importance because they no longer merely provide development and runtime infrastructures, but increasingly also create the technical and organizational framework conditions for agentic software production: context provision, access rights, tools, evaluation infrastructures, monitoring, security mechanisms, and governance.
This also shifts the significance of IT platforms: they turn from a supporting infrastructure into a potential organizational control authority of agentic software production.

\subsection{From Direct Control to Systemic Governance}

The more autonomously AI agents act, the less governance can be exercised through direct human control of individual development steps. Human intention must therefore increasingly be anchored in advance in persistent and machine-interpretable SE assets. Control thus shifts from the direct supervision of individual work steps towards the design of rules, boundaries, and verifiable objectives for agentic SE.

As a result, governance merges directly with software production: instead of documenting rules and security policies in a purely declarative manner, they are integrated directly into the development and execution process as machine-interpretable SE assets. AI agents move autonomously within these predefined guardrails, while their results in turn provide the necessary evidence for automated approvals and well-founded human decisions.

A distinction must be made here between technical governance and regulatory compliance. The EU AI Act establishes regulatory requirements for certain AI systems and their providers and deployers, but does not constitute a complete governance model for agentic SE. Production-ready agentic software systems rather require a multi-layered control regime that links human intention, requirements, and specification with AI agent actions, generated SE assets, V\&V, and the runtime behavior of the system. Traceability thus becomes a central mechanism: it must make transparent how requirements and design decisions enter into AI agent actions and generated SE assets, and how their conformance and quality are checked through V\&V.

With increasing autonomy, the relationship between production and control capacity shifts. Agentic systems can generate and modify software at a speed and volume that conventional V\&V procedures may no longer be able to keep up with. What matters, therefore, is not only the amount of generated code, but whether control capacity can keep pace with growing production capacity~\cite{cotroneo2025human}. If the ability to generate software grows faster than the ability to verify its correctness, security, maintainability, and system compatibility, V\&V can become the critical bottleneck of software production~\cite{alenezi2026rethinking}.

This decoupling is particularly relevant for security. In a controlled user study, Perry et al. showed that participants with an AI code assistant produced less secure code in security-related programming tasks and at the same time were more likely to believe they had produced secure code~\cite{perry2023users}. The risk thus arises not only from faulty or insecure AI-generated SE assets, but also from a possible overestimation of one's own ability to control them. With falling costs and increasing speed of code generation, independent and systematic V\&V therefore gains in importance.

The central systemic challenge of agentic software production thus lies in synchronizing production and control capacity. The more software production is automated and accelerated, the more powerful the mechanisms of its V\&V and governance must become. Only when automated tests, static and dynamic analyses, security checks, runtime controls, and human judgment can keep pace with the speed of agentic software production can code abundance be translated into sustainable software productivity. Agentic software production thus requires not less control, but a different form of control: less direct supervision of individual work steps and more explicit, machine-processable rules, continuous V\&V, and traceable chains of control.

\section{Path Dependencies and Design Options of Digital Self-Transformation}
\label{sec:design}

The fourth phase of digitalization is not a technologically determined development. Generative and agentic AI do create new technical possibilities, but which of these possibilities actually take effect, which actors benefit from them, and which dependencies emerge is co-determined by existing technical (see Section~\ref{sec:software}), economic (see Section~\ref{sec:marginalcosts}), organizational, and institutional (see Section~\ref{sec:dynamics}) structures.

The central design question is therefore not only how AI can be adopted as smoothly as possible. What matters, rather, is who holds the capability for the sovereign development, modification, and control of digital systems. The design task is to enable positive development paths without creating new technological and institutional lock-ins. For Germany and Europe, this means in particular building their own competencies along the software production chain and preserving the ability to choose between different technological paths.

\subsection{The Software Engineering Lottery}

Technological development does not proceed merely as a selection of respectively superior technologies. It is shaped by feedback between technologies, infrastructures, investments, competencies, and ecosystems. Arthur describes such developments as path-dependent processes in which positive feedback and self-reinforcing mechanisms can lead to the stabilization of technological paths once taken, even if alternative solutions are not fundamentally inferior~\cite{arthur1989competing}.

Hooker's concepts of the hardware lottery and the software lottery illustrate this dynamic for AI research. Research directions prevail in part because they are particularly easy to implement on the hardware and software available at the time. Technological possibilities are thus not evaluated independently of their material infrastructure, but are co-shaped by it~\cite{hooker2021hardware}. The rise of deep learning illustrates this connection: the combination of suitable algorithms, large data sets, and GPU-based computing power enabled scaling that had previously not been practicable~\cite{krizhevsky2012imagenet}. This favored a development path in which more powerful hardware, larger models, growing demand, and expanding software ecosystems increasingly reinforced one another.

Path dependencies thus exist for the fourth phase of digitalization as well: in addition to hardware, the software production infrastructure itself is increasingly becoming a factor of technological path dependency~\cite{gururaja2023build}. AI models, AI agent frameworks, development environments, tools, training data, benchmarks, engineering practices, and competencies reinforce one another. Anyone who has powerful AI coding agents, for example, but lacks suitable SE assets, verification mechanisms, and organizational competencies can exploit their potential only to a limited extent. Conversely, the widespread use of certain AI agent and platform architectures creates incentives to align development processes and competencies precisely with these systems.

For software production, this dynamic can be described as the \textbf{software engineering lottery}: not only the available hardware and software influence which AI systems can be developed successfully; the existing SE competencies, SE assets, and production environments also shape which forms of agentic software production prevail in practice and can be scaled. Technologies are thus not evaluated and deployed independently of the conditions of production, but are co-shaped by them. The ability to have software developed by AI itself becomes an infrastructural competence.

For Europe, this implies a central design option: technological sovereignty must not end with semiconductors and data centers. It must also encompass the software production chain. This includes open development environments, interoperable tool interfaces, high-quality software and engineering data, European benchmarks, reproducible evaluation procedures, and research into alternative agent and system architectures. Especially in areas where standards and ecosystems are not yet fully consolidated, there are starting points for promoting alternative development paths and limiting dependencies early on.

\subsection{Software Production Lock-In}

The economic dimension of this development is closely linked to the concentration of AI infrastructures. Frontier AI requires high investments in computing power, models, data, energy, and highly qualified personnel. Economies of scale can thereby lead to concentration on a few providers~\cite{korinek2025concentrating}. At the same time, the costs of certain forms of knowledge synthesis and software production are falling. This opposing development -- concentration of the means of production combined with falling marginal costs of individual services -- can favor an asymmetric value-creation structure in which a few providers control central production resources, while their services are used by a large number of users and organizations.

This asymmetry is particularly relevant for software production. When AI agents increasingly generate software, economic dependency no longer arises only from the use of finished software, but increasingly also from tying software production to specific technologies and infrastructures. Companies and public organizations may align their development processes, knowledge bases, tools, tests, and competencies with specific AI models, AI agent frameworks, or AI platforms. With increasing integration, this can give rise to switching costs: proprietary interfaces, specific data and context structures, adapted development processes, acquired competencies, and complementary tools can make switching to alternative providers technically complex and economically unattractive. Research on cloud computing shows that lack of interoperability and portability as well as proprietary interfaces in particular are essential mechanisms of vendor lock-in~\cite{opara2016critical,kaur2017interoperability}.

For agentic software production, this form of dependency can be conceptualized as \textbf{software production lock-in}. An organization can become dependent at different levels: through the binding of its software systems to specific platforms (system lock-in), through the binding of its technical infrastructure to specific providers (infrastructure lock-in), or through the binding of its software production processes to specific AI models and production environments (software production lock-in). The latter can arise when essential parts of an organization's own software production capability -- such as development knowledge, agent workflows, evaluation procedures, context and knowledge bases, or technical interfaces -- are tailored to the services of a particular AI model or AI platform. The higher the resulting switching costs and the lower the interoperability with alternative production environments, the stronger this dependency can become.
Agentic AI can thus move lock-in one level forward: from dependency on digital systems and infrastructures to dependency on the conditions under which digital systems are produced.

This also shifts the question of digital sovereignty from control over finished applications to control over the conditions of production of digital systems. Sovereignty here does not mean complete independence from external providers, but the ability to choose between relevant alternatives, assess dependencies, and, where necessary, change production conditions or switch providers. \textbf{Sovereign software production} must therefore preserve options at several levels: in computing infrastructures, data, software development environments, AI models, AI agent frameworks, interfaces, and evaluation procedures.

At the same time, the lowering of the production barrier must not be equated with an automatic increase in aggregate productivity. Generative and agentic AI exhibits essential characteristics of a general-purpose technology (GPT): its productive use requires complementary investments in processes, organization, business models, and human capital. Brynjolfsson, Rock, and Syverson describe this relationship as the productivity J-curve. In the introduction phase, the productivity effects of a GPT can lag behind its technical possibilities, because complementary intangible investments first have to be built up~\cite{brynjolfsson2021productivity}. The transfer of this mechanism to generative AI is also discussed in recent analyses as a relevant explanation for the delayed realization of productivity potentials~\cite{calvino2025generative}.

For digital self-\hspace{0pt}transformation, this means: the decisive productivity lever does not lie in the isolated introduction of AI coding agents. Rather, it arises from the reorganization of software production -- from requirements and specification through architecture and implementation to testing, operation, maintenance, and evolution. Agentic SE is thus not primarily a new class of tools, but a new production paradigm whose productivity potential can only be realized through complementary organizational, technical, and competence-related changes.

Empirical results already support the assumption of considerable productivity potential, but at the same time show its context dependency. Randomized field experiments with 4,867 software developers at Microsoft and Accenture, among others, found on average a 26.08\% increase in the number of completed tasks when using an AI coding assistant; at the same time, the results varied considerably between the individual experiments~\cite{cui2026effects}. The results thus demonstrate a productivity effect under the conditions studied, but cannot be directly interpreted as a proportional increase in software quality or in aggregate economic output. Precisely for this reason, SE gains in importance as a discipline of system design, quality assurance, and organizational embedding: the cheaper the production of software and its SE assets becomes, the more important becomes the ability to systematically ensure their quality, integration, operation, and long-term evolution.

\subsection{Digital Change Capability}

The technological and economic path dependencies described above imply a further prerequisite for digital self-\hspace{0pt}transformation: organizations must be able to continuously adapt and reconfigure their digital conditions of production~\cite{li2019dynamic,teece2007explicating}. This is a particular challenge, because existing IT landscapes, procurement models, role models, governance structures, and qualification profiles are themselves path-dependent. The more these structures are geared to specific technologies and modes of production, the higher the organizational and economic costs of changing them can become.

Digitalization has been and often still is organized in the form of projects: an application is introduced, a process is digitalized, or a platform is built. In the fourth phase of digitalization, this pattern alone is no longer sufficient. If software can increasingly be generated, verified, and changed dynamically, its continuous evolution itself becomes part of the digital mode of production. The organizational challenge thus shifts from the successful execution of individual digitalization projects to the lasting ability to evolve digital systems and their conditions of production.

This applies particularly to public institutions. Federal competencies, heterogeneous IT landscapes, differing procurement logics, and fragmented responsibilities can hamper the continuous evolution of digital systems. Germany's DigitalPakt Schule (Digital Pact for Schools)~\cite{digitalpaktschule2026bilanz} can serve as an example of the limits of a primarily infrastructural perspective: technical equipment creates necessary preconditions, but does not by itself create a lasting capability for digital change. Transferred to software production, this means: owning AI models generates value only once complementary organizational capabilities and governance structures have been established.

In the fourth phase of digitalization, this requirement becomes more acute. Organizations must not only operate digital systems, but increasingly organize and evolve their own digital production capability. This includes the ability to formulate precise requirements and specifications, to develop architectures, to deploy and orchestrate AI agents, to perform continuous V\&V, and to assess risks, trade-offs, and uncertainties.

Software engineering thus becomes a key organizational competence. It provides methods, models, processes, and quality-assurance mechanisms with which organizations can develop, change, and evolve their digital systems in a controlled manner. In agentic software engineering, this capability comprises not only the technical mastery of AI agents, but increasingly also the ability to assess their results independently and to decide responsibly on their use. Digital change capability thus does not mean being able to deploy every new technology, but having the ability to deliberately change one's own digital conditions of production while keeping their consequences under control.

Human judgment thereby becomes part of the technical and organizational trust infrastructure. Humans need not control every single action of an AI agent~\cite{zhu2026designing}; they must, however, remain able to define its scope of action, assess results on the basis of independent evidence, decide on trade-offs, and take responsibility for the use of the results. This at the same time broadens the understanding of V\&V: for agentic software engineering, it is not sufficient to check only the final product. Relevant development processes, agent interactions, knowledge sources used, tools, and decision and handover points must also be observable and evaluable -- insofar as required for the question at hand. Trustworthiness thus does not arise from the mere claim that AI-generated software is powerful, but from a traceable chain of trust consisting of specification, traceability, independent evidence, tests, formal or analytical proofs, and the assumption of human responsibility. Empirical studies of the deployment of software agents already show that human oversight takes different forms -- from defining scopes of action and joint planning through runtime monitoring to post-hoc review~\cite{dhanorkar2026human}.

The central resource of digital self-\hspace{0pt}transformation is therefore not computing power alone, but the ability to distinguish between generatable information, technical evidence, and justified validity. This ability must be anchored both individually, as a competence of software engineers, and institutionally, in the form of suitable processes, roles, verification mechanisms, and technical infrastructures.

\subsection{European Sovereignty}

The technological, economic, and epistemic path dependencies give rise to a central European design task. The notion of \emph{digital sovereignty} is itself not an analytically sharply defined category, but a concept that is controversially debated and heterogeneously filled in research and politics~\cite{broeders2023search,fratini2024digital}. It oscillates between a descriptive understanding as the capability to control critical digital infrastructures and a normative understanding as a political guiding principle of European self-assertion~\cite{floridi2020fight,santaniello2025attributes}. Digital sovereignty should therefore not be misunderstood as complete autarky. Rather, it denotes the ability to understand, assess, shape, and, where necessary, change critical digital systems and their conditions of production.

This ability cannot be taken for granted. Comparative analyses of European digital policy show that the political sovereignty discourse is not followed by corresponding substantive policy change in all policy fields, and that European AI policy in particular is shaped more by competition-oriented objectives in international systemic rivalry than by a fundamental alternative to the logic of global competition~\cite{falkner2024digital}. The national sovereignty discourse itself is not homogeneous either, but an expression of different political interests and interpretive patterns~\cite{lambach2022narratives}. A scientifically grounded examination of European sovereignty in software production must reflect this ambivalence: sovereignty policy can effectively reduce dependencies, but it can also function as a political legitimation narrative without being followed by corresponding structural investments.

For digital self-\hspace{0pt}transformation, the production-oriented notion of sovereignty advocated here implies a shift in focus from finished applications to the upstream conditions of production of digital systems. It is not enough to build up European computing capacity, cloud infrastructure, or AI models. Europe must also have the ability to master the software production that builds on them. Following the distinction between system, infrastructure, and software production lock-in introduced in Section~\ref{sec:design}, the central fields of action can be arranged according to the three levels at which dependencies can arise:

\begin{itemize}
\item at the \textbf{infrastructure level}: powerful and interoperable development and AI agent infrastructures as well as secure and auditable engineering and agent platforms;
\item at the \textbf{system level}: open standards and interfaces for agentic SE as well as open digital commons and open-source ecosystems;
\item at the \textbf{software production level} in the narrower sense: high-quality data and knowledge bases for SE, European evaluation, testing, and V\&V infrastructures, research into alternative AI, agent, and computer architectures, and above all the education and continuous development of the corresponding engineering competencies.
\end{itemize}

That dependencies in the AI era fundamentally extend beyond classical portability problems is also evident beyond the European context: current analyses of the procurement of AI systems point out that AI-specific lock-in mechanisms are no longer confined to software interfaces, but increasingly concern intellectual property and algorithmic competencies, since AI models trained on proprietary platforms are sometimes inseparably tied to their training environment~\cite{itea2026vendorlockin}. From a different perspective, this confirms the thesis advanced here that agentic AI tends to move lock-in one production level forward.

The German Informatics Society (Gesellschaft für Informatik, GI) accordingly classifies AI-based SE as a key technology for Germany's technological innovative strength and digital sovereignty~\cite{gi2026aikbse}. Since June 2026, this classification has also had a direct political counterpart: with the \emph{European Technological Sovereignty Package}, the European Commission has for the first time presented a policy framework that explicitly treats semiconductors, cloud and AI infrastructure, and software as interconnected, mutually reinforcing elements of a common sovereignty strategy~\cite{eucom2026techsovereignty}. It is noteworthy that the accompanying EU Open Source Strategy for the first time anchors open source at the highest political level not merely as a technical tool but as a strategic instrument of European sovereignty -- a clear indication of the practical relevance of the system level outlined above. This prioritization has also been consolidated at the intergovernmental level, for instance at the Franco-German Summit on European Digital Sovereignty in November 2025~\cite{eucouncil2025summit}. In addition, recent policy-oriented studies propose, with the concept of a European EuroStack, a multi-layered reference model that conceives of data centers, cloud, data, AI models, and applications as a sovereignty architecture to be designed jointly~\cite{bria2025eurostack}.

For Germany and Europe, this calls for a strategic change of perspective. What should be promoted is not only the development of individual AI models, but a powerful European software engineering ecosystem for the AI age. Research, industry, the public sector, and universities need shared infrastructures in which agentic development methods can be developed, tested, evaluated, certified, and transferred into real-world applications. Whether the current political momentum actually leads to a sustainable reduction of structural dependencies or remains largely declaratory is -- in line with the criticism cited above -- an open empirical question.

The decisive design option therefore consists in using the current phase of high technological dynamics in research~\cite{SRIA2026_Matrix,abrahao2025software,ahmed2025artificial,hassan2026agentic}, industry, and society before new lock-ins become entrenched. Europe does not have to develop every technological path itself. It must, however, preserve the ability to choose between paths, to bring forth its own alternatives, and to control critical dependencies.

\section{Summary and Outlook}
\label{sec:summary}

Even though the fourth phase of digitalization is not complete but a currently emerging constellation, its features and structural breaks are already recognizable:
The fourth phase of digitalization differs qualitatively from the preceding phases. Digitalization no longer merely transforms information, processes, organizations, and value-creation models. It is beginning to transform its own production base. Digital systems are increasingly developed, tested, operated, and evolved with the help of digital systems. Digitalization thus becomes recursive: it becomes digital self-\hspace{0pt}transformation.

The central technical mechanism of this development is the transition from generative AI to agentic software production. AI systems no longer merely generate individual code fragments, but can increasingly pursue coherent engineering tasks, operate development environments, modify software, run tests, and iteratively improve results. The boundary of automation thus shifts from code generation to parts of the software engineering process.

This development changes the division of labor between humans and machines, but it does not make SE obsolete. On the contrary: the more implementation is automated, the more important become the tasks that determine the purpose, quality, and limits of the system. System intention, architecture, quality goals, V\&V, security, traceability, and accountability become the central objects of engineering design.

The economic consequence is a new structure of scarcity. The marginal costs of certain forms of software creation are falling, which can considerably expand the space of economically viable software. At the same time, system understanding, robust requirements, high-quality data, computing and energy infrastructure, independent evidence, and human judgment remain scarce. Moreover, the potential productivity gains from AI are not realized automatically through the introduction of individual tools, but through complementary investments in organization, processes, competencies, and trust infrastructures.

Agentic SE thereby becomes the key technology of the fourth phase of digitalization. It is the technology with which the digital production base itself is being redesigned. At the same time, SE becomes a key competence for companies, the state, and society: whoever can master the creation and modification of digital systems possesses a central prerequisite for digital value creation, technological innovation capability, and digital sovereignty.

This is of particular importance for Germany and Europe. The decisive strategic question is not whether Europe can develop every foundation model or every platform itself. What matters is whether Europe has the ability to understand, design, verify, and evolve digital systems independently enough to control critical dependencies and to open up its own development paths.

The competencies required for this go far beyond prompting or the use of AI coding assistants. Next-generation software engineers must be able to model complex socio-technical systems, specify requirements precisely, design architectures, orchestrate AI agents, generate evidence for the quality of their results, and recognize the limits of machine decisions. They thus become less manual producers of code than designers and managers of digital production systems.

This results in the following desiderata for future research: SE research must develop models for human--agent collaboration, agentic development processes, multi-agent systems, machine-interpretable specifications, continuous V\&V, and trust infrastructures. Education must systematically teach these capabilities, combining a solid foundation in computer science and classical SE with new competencies for agentic systems.

At the same time, policy-makers must support the emergence of a European engineering ecosystem. This includes open standards and digital commons, secure and interoperable agent infrastructures, European research and testing infrastructures, as well as long-term programs for research, education, and transfer.

The central question of the fourth phase of digitalization is therefore not how much software AI can generate. Rather, it is whether people, organizations, and societies preserve and expand the ability to deliberately determine how digital systems come into being. This is precisely where the strategic significance of SE lies.

When digitalization enters a phase of digital self-\hspace{0pt}transformation, SE turns from an important engineering discipline into a prerequisite for being able to shape digital transformation itself. Agentic SE thus goes beyond a new form of software development: it is one of the central socio-technical prerequisites for the digital sovereignty of Germany and Europe. Whether and to what extent these possibilities are actually translated into structural independence, however, depends on institutional and political framework conditions.

\bibliographystyle{spmpsci}
\bibliography{author}

\selectlanguage{ngerman}

\title*{Wenn Digitalisierung sich selbst verändert: \newline KI, Software und die nächste technische Ordnung}
\titlerunning{Digitale Selbsttransformation}
\author{Ina K. Schieferdecker}
\institute{Technische Universität Berlin \at Einsteinufer 25, 10587 Berlin, \email{ina.schieferdecker@tu-berlin.de}}

\maketitle

\abstract{Agentische Künstliche Intelligenz (KI) markiert eine neue Phase der Digitalisierung: Digitalisierung beginnt, auf ihre eigene technische Produktionsbasis zurückzuwirken. Während frühere Phasen darauf ausgerichtet waren, analoge Informationen zu digitalisieren, Prozesse zu automatisieren und digitale Wertschöpfungsnetzwerke aufzubauen, übernimmt KI zunehmend Aufgaben bei der Entwicklung der digitalen Systeme selbst. Insbesondere im Software Engineering (SE) können agentische KI-Systeme bereits heute Entwicklungsaufgaben über mehrere Schritte hinweg planen, ausführen, prüfen und iterativ weiterentwickeln.
Damit wird Digitalisierung rekursiv: Digitale Systeme unterstützen nicht mehr nur die Digitalisierung anderer Branchen, sondern werden zunehmend selbst durch digitale Systeme entwickelt und weiterentwickelt. Die technische Produktionsbasis der Digitalisierung wird damit zum Gegenstand ihrer eigenen Transformation. Dies verändert nicht nur die Arbeitsteilung zwischen Mensch und Maschine im SE, sondern auch die Rolle des SE insgesamt. Das Kapitel analysiert diesen soziotechnischen Wandel und argumentiert, dass SE sich von einer primär unterstützenden Ingenieurdisziplin zu einer zentralen Infrastruktur gesellschaftlicher Wertschöpfung sowie technologischer und digitaler Souveränität entwickelt.}

\medskip
\noindent\textit{Hinweis: Dies ist die deutsche Originalfassung. Die englische Übersetzung bildet den ersten Teil dieses Dokuments.}

\section{Einleitung}
\label{de:sec:Einleitung}

Wenn Künstliche Intelligenz (KI) beginnt, nicht mehr nur Branchen zu digitalisieren, sondern die eigene technische Produktionsbasis digitalisiert und auf sie zurückwirkt, betritt die Digitalisierung eine neue, rekursive Entwicklungsstufe: die \textbf{digitale Selbsttransformation}. Während frühere Phasen darauf ausgerichtet waren, analoge Informationen zu digitalisieren und Prozesse zu automatisieren, übernimmt agentische KI zunehmend Aufgaben bei der Entstehung, Prüfung und Evolution der digitalen Systeme selbst. Software wird damit vom Instrument der Digitalisierung zu deren Produkt – und das Software Engineering (SE) vom unterstützenden Handwerk zur strategischen Infrastruktur gesellschaftlicher Wertschöpfung und technologischer Souveränität.

Damit verschiebt sich der Gegenstand der Digitalisierung. Sie beschränkt sich nicht länger nur auf Informationen, Prozesse und Wertschöpfungsstrukturen, sondern erweitert sich auf die Prozesse, durch die digitale Systeme selbst entstehen, geprüft und verändert werden. Diese \textbf{digitale Selbsttransformation} geht über eine weitere Stufe der Automatisierung von Softwareentwicklung hinaus. Sie verändert die Produktionslogik von Software, die Arbeitsteilung zwischen Menschen und Maschinen sowie die Anforderungen an Qualitätssicherung, Verantwortung und Governance. Zugleich entstehen neue ökonomische und technologische Pfadabhängigkeiten: Je stärker Software zur Grundlage digitaler Wertschöpfung wird und je stärker ihre Herstellung durch KI-basierte Systeme geprägt wird, desto strategischer wird die Kontrolle über die technischen, infrastrukturellen und wissensbasierten Bedingungen ihrer Produktion. Die Fähigkeit, Software zu entwickeln, zu prüfen und weiterzuentwickeln, wird damit selbst zu einer Frage technologischer Handlungsfähigkeit.

Im Zentrum dieser Entwicklung steht daher nicht allein die Leistungsfähigkeit generativer oder agentischer KI, sondern die Frage, wie Software Engineering (SE) unter diesen Bedingungen neu gestaltet werden muss. Wenn Maschinen zunehmend Implementierungs- und andere Entwicklungsaufgaben übernehmen, verschiebt sich die ingenieurmäßige Wertschöpfung von der manuellen Erzeugung von Code hin zur Präzisierung von Systemintentionen, zur Architektur komplexer Systeme, zur Definition und Durchsetzung von Qualitätszielen sowie zu Verifikation und Validierung (V\&V) sowie Governance. SE wird damit in zweifacher Hinsicht zum Gegenstand der Digitalisierung: Zum einen werden seine Prozesse und Tätigkeiten selbst zunehmend durch KI-Systeme verändert; zum anderen wird SE zur zentralen Fähigkeit, die entstehenden digitalen Produktionssysteme zu entwerfen, zu kontrollieren und weiterzuentwickeln. Damit gewinnt SE eine strategische Bedeutung, die über seine traditionelle Rolle als unterstützende Ingenieurdisziplin hinausgeht.

Aus dieser Perspektive untersucht das Kapitel die \textbf{vierte Digitalisierungsphase} als Prozess einer digitalen Selbsttransformation. Es betrachtet zunächst die technischen und ökonomischen Grundlagen der sich herausbildenden Softwareproduktionsweise in Abschnitt~\ref{de:sec:Digitalisierung}, analysiert anschließend die Veränderungen von Softwareproduktion und SE in Abschnitt~\ref{de:sec:Software} und untersucht anschließend die Veränderung der kognitiven Grenzkosten der Softwareproduktion in Abschnitt~\ref{de:sec:Grenzkosten} sowie die soziotechnischen Dynamiken der Softwareproduktion (Abschnitt~\ref{de:sec:Dynamiken}) und ökonomischen Pfadabhängigkeiten sowie mögliche Gestaltungsoptionen in Abschnitt~\ref{de:sec:Gestaltung}. Im Mittelpunkt steht die Frage, welche Rolle agentisches SE für die zukünftige digitale Wertschöpfung einnimmt und welche Bedeutung ihm für die technologische und digitale Souveränität Deutschlands und Europas zukommt. Zusammenfassung und Ausblick schließen das Kapitel in Abschnitt~\ref{de:sec:Zusammenfassung} ab.

\section{Die Phasen der Digitalisierung}
\label{de:sec:Digitalisierung}

Die Digitalisierung hat ihren Gegenstandsbereich über mehrere Phasen hinweg kontinuierlich erweitert (siehe Tabelle~\ref{de:tab:Phasen}\footnote{Zur Erläuterung der letzten Spalte Grenzkosten in dieser Tabelle, siehe Abschnitt~\ref{de:sec:Grenzkosten}.}). Auf die Digitalisierung analoger Informationen folgten die Digitalisierung und Automatisierung von Prozessen sowie die digitale Vernetzung und Transformation von Wertschöpfungsstrukturen. Mit dem Aufkommen generativer und insbesondere agentischer Künstlicher Intelligenz (KI) digitalisiert sich Digitalisierung zunehmend rekursiv selber. 

Die ersten drei Phasen der Digitalisierung bilden den technologischen und ökonomischen Entwicklungspfad, vor dessen Hintergrund diese neue Qualität sichtbar wird. In der ersten Phase der technischen Digitalisierung wurden analoge Informationen in digitale Repräsentationen überführt. Mit Beginn des Computerzeitalters wurden beispielsweise Papierakten, Buchungslisten oder andere analoge Aufzeichnungen in Datenstrukturen überführt, die durch Software verarbeitet werden konnten. Im Zentrum stand die digitale Repräsentation bestehender Strukturen.

In der zweiten Phase der Prozess-Digitalisierung wurde der Gegenstandsbereich auf Arbeits- und Geschäftsprozesse erweitert. Software und digitale Vernetzung ermöglichten deren Automatisierung und Optimierung, beispielsweise durch E-Mail, ERP-Systeme oder Online-Shops. Während sich dadurch die Durchführung von Prozessen grundlegend verändern konnte, blieben die zugrunde liegenden Geschäftsmodelle vielfach weitgehend erhalten.

Die dritte Phase, die digitale Transformation, erfasste schließlich nicht mehr nur einzelne Prozesse, sondern zunehmend ganze Wertschöpfungsstrukturen. Plattformen, Cloud Computing und das Internet der Dinge ermöglichten neue Formen der Vernetzung, Koordination und Skalierung. Dadurch entstanden neue Geschäftsmodelle und Organisationsformen. Die Digitalisierung wurde damit selbst zu einem zentralen Treiber wirtschaftlicher und gesellschaftlicher Veränderung.

Das charakteristische Merkmal der vierten Phase ist die Rekursivität der Digitalisierung, die digitale Selbsttransformation: Die technischen Systeme und Prozesse, mit denen Digitalisierung realisiert wird, werden selbst zum Gegenstand digitaler Transformation. Insbesondere agentische KI ermöglicht dabei Entwicklungs- und Optimierungsprozesse, in denen digitale Systeme Ziele interpretieren, Teilaufgaben planen, Werkzeuge nutzen, Ergebnisse bewerten und auf Grundlage von Rückkopplungen weitere Aktionen ausführen. Der qualitative Bruch der vierten Phase liegt in ihrer inhärenten Rekursivität: Anstatt ‚nur‘ externe Branchen zu durchdringen, digitalisiert die Digitalisierung ihre eigene Genese. In den Fokus rücken jene maschinellen und menschlichen Prozesse, die Software erschaffen und wandeln. Software transzendiert damit ihre Rolle als bloßes Werkzeug der digitalen Transformation und wird zum eigenständigen, agentisch erzeugten Resultat einer sich selbst transformierenden Produktionsbasis.

\begin{table}
\centering
\renewcommand{\arraystretch}{1.2}
\setlength{\tabcolsep}{2pt}
\begin{tabular}{|>{\raggedright\arraybackslash}p{0.8cm}|>{\raggedright\arraybackslash}p{1.8cm}|>{\raggedright\arraybackslash}p{1.6cm}|>{\raggedright\arraybackslash}p{2.1cm}|>{\raggedright\arraybackslash}p{2.5cm}|>{\raggedright\arraybackslash}p{2cm}|}
\hline
    \textbf{Phase} & 
    \textbf{Bezeichnung} & 
    \textbf{Wann} & 
    \textbf{Gegenstand} & 
    \textbf{Charakteristik} & 
    \textbf{Sinkende Grenzkosten} \\
\hline
    1 & 
    Technische Digitalisierung (Digitization) & 
    1960--1990er & 
    Informationen & 
    Analoge Informationen werden digital repräsentiert und verarbeitet. &
    Speicherung und Verarbeitung \\
\hline
    2 & 
    Prozess-Digitalisierung (Digitalization) & 
    1990--2010er & 
    Prozesse & 
    Digitale Prozesse werden automatisiert und optimiert. &
    Prozess-ausführung \\
\hline
    3 & 
    Digitale\newline Transformation & 
    2010--2020er & 
    Wertschöpfungs-strukturen & 
    Vernetzte digitale Prozesse ermöglichen neue Geschäftsmodelle und Organisationsformen. &
    Kommunikation und Koordination \\
\hline
    4 & 
    Digitale\newline Selbsttrans-formation & 
    seit 2020er & 
    Technische\newline Produktionsbasis der Digitalisierung & 
    Digitale Systeme übernehmen zunehmend die Entwicklung, Prüfung, Anpassung und Optimierung digitaler Systeme selbst. &
    Kognitive Synthese \\
\hline
\end{tabular}
\caption{Die Phasen der Digitalisierung}
\label{de:tab:Phasen}
\end{table}

Die vierte Phase lässt sich besonders deutlich am Verhältnis von Produkten, Prozessen und ihrer Produktion beobachten. In der Medienindustrie beispielsweise führte die Digitalisierung zunächst zur Entstehung digitaler Trägermedien, anschließend zur Digitalisierung von Vertrieb und Geschäftsprozessen und schließlich zu plattformbasierten und datengetriebenen Geschäftsmodellen. Mit generativer KI entsteht nun eine weitere Qualität: Inhalte können zunehmend situativ und individuell erzeugt werden. Die Grenze zwischen Produktion und Rezeption verschiebt sich, weil digitale Inhalte nicht mehr ausschließlich vorab produziert und anschließend genutzt werden, sondern teilweise erst im Prozess ihrer Nutzung entstehen.

Ein vergleichbarer Wandel zeigt sich in der Industrieautomatisierung. Während frühe Automatisierung auf deterministisch programmierten Abläufen beruhte, ermöglichten vernetzte Systeme zunächst die Integration von Produktions- und Geschäftsprozessen. Digitale Zwillinge, Internet of Things und Predictive Maintenance erweiterten diese Fähigkeiten um datenbasierte Analyse und Optimierung. Die Optimierungsziele und Entscheidungsregeln blieben jedoch grundsätzlich durch Menschen vorgegeben. In der vierten Phase verschiebt sich diese Grenze: KI-basierte Systeme können zunehmend Produktionszustände interpretieren, Handlungsoptionen erzeugen, deren Ergebnisse bewerten und ihre nächsten Aktionen auf dieser Grundlage bestimmen. Entscheidend ist dabei nicht die vollständige Autonomie einzelner Maschinen, sondern die Entstehung rekursiver Feedbackschleifen, in denen digitale Systeme an der Veränderung ihrer eigenen Prozesse beteiligt sind.

\subsection{Die digitale Selbsttransformation}
\label{de:sec:Selbsttransformation}

Das charakteristische Merkmal der vierten Digitalisierungsphase ist die Rückwirkung der Digitalisierung auf ihre eigene technische Produktionsbasis (siehe Tabelle~\ref{de:tab:Selbsttransformation}). Agentische KI ist ein wesentlicher technologischer Enabler dieser Entwicklung. Im SE manifestiert sie sich als agentisches SE, in dem KI-Agenten zunehmend mehrstufige Entwicklungsaufgaben über den Softwarelebenszyklus hinweg übernehmen und ihre Handlungen auf der Grundlage von Rückkopplungen koordinieren. Die entscheidende Veränderung liegt daher nicht allein darin, dass KI Code erzeugt, sondern darin, dass sie zunehmend an den Prozessen beteiligt wird, durch die Software geplant, entwickelt, geprüft und weiterentwickelt wird.

\begin{table}
\centering
\renewcommand{\arraystretch}{1.2}
\setlength{\tabcolsep}{2pt}
\begin{tabular}{|p{3.3cm}|>{\raggedright\arraybackslash}p{8.5cm}|}
\hline
    \textbf{Merkmal} & 
    \textbf{Digitale Selbsttransformation} \\
\hline
    Gegenstand & 
    Entwicklung, Anpassung und Optimierung digitaler Systeme selbst\\
\hline
    Technologische Basis & 
    KI-Modelle, agentische KI, KI-augmentierte SE Assets sowie Trainings-, Simulations- und Evaluationsumgebungen\\
\hline
    Produktionsweise 
    & Rekursive, feedbackbasierte und zunehmend agentisch ausgeführte Entwicklungsprozesse\\
\hline
    Produkte und Dienste 
    & Adaptive, individualisierte, selbstoptimierende und zunehmend autonome Systeme\\
\hline
    Organisation & 
    Mensch-KI- und Multi-Agenten-Kollaboration\\
\hline
    Wertschöpfung & 
    Kognitive Industrialisierung und hochgradige Individualisierung\\
\hline
    Governance & 
    Kontinuierliche V\&V Überwachung und Steuerung\\
\hline
Metriken & Qualität, Robustheit, Sicherheit, Nachhaltigkeit, Vertrauenswürdigkeit sowie Kosten und Produktivität\\
\hline
\end{tabular}
\caption{Charakteristika der vierten Digitalisierungsphase: Digitale Selbsttransformation}
\label{de:tab:Selbsttransformation}
\end{table}

Diese Analyse knüpft an die von Verhoef u.a.~\cite{verhoef2021digital} vorgenommene Unterscheidung von Digitization, Digitalization und Digital Transformation an und erweitert diese Perspektive um die vierte, gegenwärtig entstehende Phase. Während Verhoef u.a.~\cite{verhoef2021digital} die drei Phasen insbesondere aus der Perspektive der Transformation von Unternehmen, Ressourcen, Organisationsstrukturen und Geschäftsmodellen betrachten, verschiebt sich in der vierten Phase der Gegenstand der Digitalisierung selbst.

Für die folgende Betrachtung wird der Begriff des \textbf{SE Assets} genutzt, um einerseits auf den Unterschied zwischen klassischen SE-Artefakten und agentischen Assets im SE zu verweisen. Klassische SE-Artefakte wie Anforderungsdokumente, Architekturdiagramme, Code oder Testfälle sind typischerweise persistente, versionierte und von Menschen autorisierte Ergebnisse eines Arbeitsschritts im SE~\cite{SWEBOK2024}. Im agentischen SE entstehen dagegen laufend Zwischenprodukte wie Agenten-Pläne, Kontext- und Memory-Objekte oder probeweise erzeugte und wieder verworfene Code-Versionen, die maschinell erzeugt, iterativ verändert oder überschrieben werden und teilweise flüchtig sind~\cite{hassan2026agentic}. Der Begriff wird zudem verwendet, um sowohl KI-erzeugte~\cite{mandl2026ai} als auch KI-genutzte Zwischen- und Endprodukte zu bezeichnen. Er ermöglicht damit eine kompakte Beschreibung wesentlicher Dynamiken des SE in der digitalen Selbsttransformation. Der Begriff des SE Assets ist in Abgrenzung zum AI-augmented SE in~\cite{Schieferdecker2026_AI4SE} sowie zu KI-generierten Artefakten in~\cite{mandl2026ai} gewählt.

Auch wenn sich die Charakteristika der vierten Digitalisierungsphase erst im weiteren Verlauf ihrer Entwicklung vollständig bestimmen lassen, zeichnen sich wesentliche Merkmale bereits heute ab. Eine zentrale Veränderung betrifft die Produktionsweise von Software: Foundation Models (kurz KI-Modelle), SE Assets sowie Trainings-, Simulations- und Evaluationsumgebungen bilden eine neue Ressourcen- und Produktionsbasis. Im SE verdichtet sich diese Basis zu einer Konstellation aus Big Data, Big Code und Big (Software) Models: Spezifikationen, Modelle und Code sind darin nicht mehr nur Ergebnisse der Softwareentwicklung, sondern werden selbst zu maschineninterpretierbaren Ressourcen für deren Analyse, Erzeugung und Weiterentwicklung~\cite{Schieferdecker2026_AI4SE}.

Beispielsweise werden Spezifikationen mit dem gegenwärtig entstehenden sogenannten Spec-Driven Development zunehmend zu  Steuerungs- und Referenzpunkten für KI-Agenten, die daraus Pläne, Implementierungen und Tests ableiten und Ergebnisse gegen spezifizierte Eigenschaften verifizieren können~\cite{gundala2026SDD}. Im Konzept des sogenannten Loop Engineerings wird dieser Ansatz zu kontinuierlichen agentischen Feedbackschleifen erweitert, in denen KI-Agenten planen, handeln, Ergebnisse beobachten und verifizieren sowie ihre nächsten Aktionen auf dieser Grundlage bestimmen~\cite{hassan2026agentic}. Damit können KI-Agenten zunehmend Aufgaben über mehrere Phasen des Softwarelebenszyklus hinweg übernehmen und dabei selbst neue SE Assets erzeugen. Sie treten damit zunehmend aus ihrer Rolle als Konsumenten und Optimierer von SE Assets heraus und werden selbst zu deren Produzenten.

Die daraus entstehenden Organisationsstrukturen sind soziotechnisch geprägt. Menschen definieren zunehmend Ziele, Randbedingungen, Qualitätsanforderungen und Verantwortlichkeiten, während KI-Agenten Planung, Implementierung, Test, V\&V und Optimierung zumindest teilweise übernehmen. Die zentrale ingenieurmäßige Aufgabe verschiebt sich damit von der manuellen Erzeugung einzelner SE Assets hin zur Gestaltung und Kontrolle der Prozesse, in denen Menschen und Maschinen gemeinsam Software erzeugen und weiterentwickeln.

Auch die ökonomische Logik verändert sich. Die Wertschöpfung zielt zunehmend auf hochgradig individualisierte, adaptive und kontinuierlich optimierte Produkte und Dienste. Gleichzeitig verschiebt sich der Schwerpunkt der Optimierung von Kosten und Produktivität hin zu einem multidimensionalen Zusammenspiel von Kosten, Qualität, Sicherheit, Nachhaltigkeit und Vertrauenswürdigkeit. Je stärker Systeme ihre eigenen Entwicklungs- und Optimierungsprozesse beeinflussen können, desto wichtiger wird daher die Festlegung dessen, was optimiert werden soll. Die Definition von Zielen, Randbedingungen und akzeptablen Kompromissen wird zu einer zentralen technischen, organisatorischen und gesellschaftlichen Gestaltungsaufgabe.

Die digitale Selbsttransformation bezeichnet damit keine abgeschlossene Digitalisierungsphase, sondern eine gegenwärtig entstehende Konstellation, deren Reichweite und Grenzen noch offen sind. Ihr entscheidendes Merkmal ist die Rekursivität: Digitalisierung wird nicht nur weiter vorangetrieben, sondern verändert zunehmend die Bedingungen, unter denen Digitalisierung selbst produziert wird. 

\section{Die Softwareproduktion im Wandel}
\label{de:sec:Software}

Software ist der zentrale technische Träger der Digitalisierung. Fast alle digitalen Produkte, Dienstleistungen und Prozesse beruhen auf Software. Ihre Herstellung ist jedoch weit mehr als das Schreiben von Code. Bereits mit den ersten komplexen Softwaresystemen zeigte sich, dass Anforderungen systematisch erfasst, Lösungsstrukturen entworfen, Implementierungen erstellt, Qualität überprüft und Änderungen über den Lebenszyklus hinweg kontrolliert werden müssen. Mit zunehmender Komplexität und Kritikalität softwarebasierter Systeme wurde daher die Fähigkeit entscheidend, diese unterschiedlichen Tätigkeiten systematisch miteinander zu verbinden. Softwareproduktion entwickelte sich von einer primär individuellen Programmieraktivität zu einem komplexen Engineering-Prozess.

\begin{figure}
    \centering
    \includegraphics[width=\linewidth]{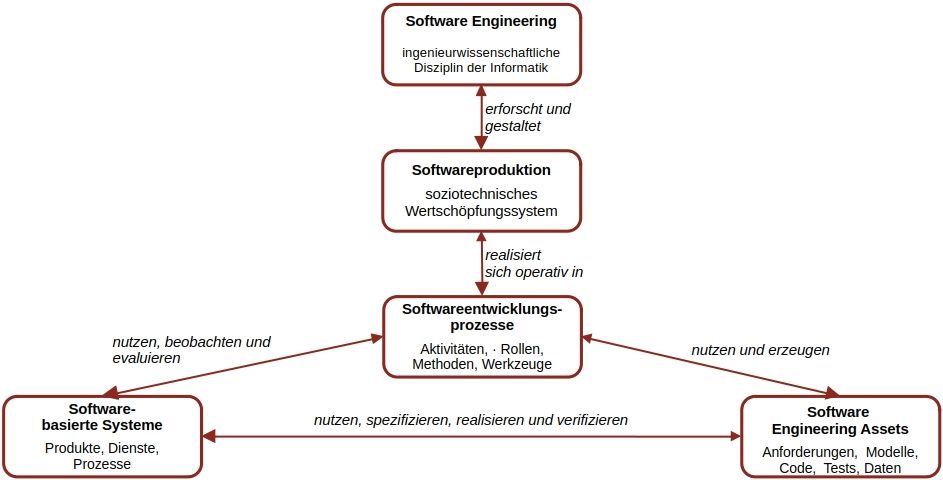}
    \caption{Beziehungen zwischen SE, Softwareproduktion, Softwareentwicklungsprozessen, SE Assets sowie softwarebasierten Systemen}
    \label{de:fig:SoftwareEngineering}
\end{figure}

Diese Entwicklung führte im Kontext der Softwarekrise der späten 1960er Jahre~\cite{nato1969software} zur Herausbildung des SE als eigenständiger Disziplin der Informatik~\cite{SWEBOK2024}. SE bezeichnet die systematische, disziplinierte und quantifizierbare Anwendung ingenieurwissenschaftlicher Prinzipien auf die Spezifikation, Entwicklung, Evolution und den Betrieb softwarebasierter Systeme~\cite{IEEE610121990}. Ziel ist es, Software nicht nur funktionsfähig, sondern auch zuverlässig, sicher, wartbar, nachvollziehbar und wirtschaftlich entwickelbar zu machen.

SE organisiert dazu den gesamten Entwicklungsprozess: von der Definition von Zielen und Anforderungen über Architektur und Implementierung bis hin zu Test, V\&V, Betrieb und Weiterentwicklung. Dabei entstehen und werden unterschiedliche SE Assets verwendet. Dazu gehören beispielsweise Anforderungen, Spezifikationen und Modelle, Architekturen, Code, Tests, Daten sowie Logs, Traces und Dokumentationen. Sie sind die strukturierten Wissens- und Arbeitsgrundlagen, auf denen die Entwicklung, Qualitätssicherung und Evolution softwarebasierter Systeme aufbauen (siehe Abbildung~\ref{de:fig:SoftwareEngineering}).

Mit der zunehmenden Bedeutung softwarebasierter Systeme wird damit nicht nur Software selbst zu einer zentralen gesellschaftlichen und wirtschaftlichen Infrastruktur. Auch die Fähigkeit, Software systematisch zu spezifizieren, zu entwickeln, zu verifizieren und weiterzuentwickeln, wird zu einer strategischen Fähigkeit. SE gestaltet und beherrscht diese Softwareproduktion und entwickelt die hierfür erforderlichen Methoden, Prozesse und Werkzeuge.

Genau hier setzt die vierte Digitalisierungsphase an (siehe Tabelle~\ref{de:tab:Assets}): Während bislang in der digitalen Transformation andere Branchen digitalisiert wurden, wird nun zunehmend auch die Branche der Softwareproduktion selbst zum Gegenstand der Digitalisierung.

Diese Veränderung betrifft mehr als die automatisierte Erzeugung von Programmcode. Im klassischen SE entstehen SE Assets überwiegend durch menschliche Ingenieurstätigkeit. Anforderungen werden formuliert, Architekturen entworfen, Implementierungen erstellt und Tests entwickelt. Automatisierungsmethoden wie Model-Driven SE~\cite{Schieferdecker2025} unterstützen dabei einzelne Tätigkeiten, die grundlegenden Arbeiten im Entwicklungsprozess bleiben jedoch überwiegend menschlich bestimmt.

Im agentischen SE verändert sich die Arbeitsteilung. KI-Agenten können mehrere Engineering-Aktivitäten miteinander verbinden: Sie können Anforderungen analysieren, Pläne erzeugen, Entwicklungswerkzeuge aufrufen, Code generieren und verändern, Tests ausführen, Ergebnisse auswerten und auf dieser Grundlage weitere Entwicklungsschritte initiieren. SE Assets sind damit nicht mehr nur Ergebnisse der Softwareentwicklung, sondern können selbst zu dynamischen Ressourcen werden, die von KI-Agenten erzeugt, interpretiert, verändert und rekursiv für weitere Entwicklungsschritte genutzt werden.

\begin{table}
    \centering
    \renewcommand{\arraystretch}{1.2}
    \setlength{\tabcolsep}{2pt}
    \begin{tabular}{|>{\raggedright\arraybackslash}p{2cm}|
                    >{\raggedright\arraybackslash}p{3cm}|
                    >{\raggedright\arraybackslash}p{3cm}|
                    >{\raggedright\arraybackslash}p{3cm}|}
        \hline
        \textbf{SE Asset} & \textbf{Klassisches SE} & \textbf{Agentisches SE} & \textbf{Agentic Loop}\\
        \hline
        Anforderungen &
        Zieldefinition &
        Agentensteuerung &
        Zieladaption\\
        \hline
        Spezifikationen / Modelle &
        Systembeschreibung &
        Agentenkontext und -steuerung &
        Modellgenerierung und -evolution\\
        \hline
        Projektpläne / Aufgaben &
        Arbeitsstrukturierung &
        Agentenplanung und -orchestrierung &
        Dynamische Aufgabenanpassung\\
        \hline
        Code &
        Implementierung &
        Generierung und Analyse &
        Kontinuierliche Weiterentwicklung\\
        \hline
        Tests &
        V\&V &
        Feedback und Evidenz &
        Automatische Testgenerierung und -optimierung\\
        \hline
        Daten &
        Informations- und Wissensbasis &
        Kontext und Evidenz &
        Kontinuierliche Anreicherung\\
        \hline
        Logs / Traces &
        Diagnose und Nachvollziehbarkeit &
        Erfahrungs- und Evaluationsdaten &
        Lern- und Optimierungsgrundlage\\
        \hline
        Policies / Engineering Guidelines &
        Richtlinien &
        Agentenvorgaben &
        Deklarative Prozesssteuerung\\
        \hline
    \end{tabular}
    \caption{SE Assets im Übergang vom klassischen zum agentischen SE -- vom statischen zum dynamischen Artefakt}
    \label{de:tab:Assets}
\end{table}

Die Veränderung lässt sich anhand der unterschiedlichen Rollen der SE Assets verdeutlichen. Anforderungen und daraus resultierende Spezifikationen und Modelle dienen im klassischen SE vor allem dazu, Ziele und Eigenschaften eines Systems für Menschen und Entwicklungswerkzeuge zu beschreiben. Im agentischen SE werden sie zunehmend zu maschineninterpretierbaren Steuerungs- und Referenzpunkten: KI-Agenten können daraus Projektpläne ableiten, Implementierungen erzeugen und Ergebnisse gegen vorgegebene Eigenschaften prüfen. Beim Spec-Driven Development (siehe Abschnitt~\ref{de:sec:Digitalisierung}) erhalten Spezifikationen in agentischen SE-Prozessen eine neue Bedeutung: Sie beschreiben nicht nur, was ein System leisten soll, sondern dienen zugleich als persistent verfügbarer Kontext und als Referenz für die Bewertung der von agentischen KI-Systemen erzeugten SE Assets.

Code verändert ebenfalls seine Rolle. Er ist nicht mehr ausschließlich das Ergebnis menschlicher Implementierung, sondern kann von KI-Agenten erzeugt, analysiert, getestet und verändert werden. Damit wird Code zunehmend zu einem Zwischenprodukt innerhalb eines iterativen Engineering-Prozesses. Tests und Evaluationen werden dabei von nachgelagerten Qualitätssicherungsaktivitäten zu kontinuierlichen Quellen von Evidenz und Rückkopplung, auf deren Grundlage KI-Agenten nachfolgende Entwicklungsentscheidungen treffen und ihre Ergebnisse iterativ weiterentwickeln.

Diese Rückkopplung wird im sogenannten Loop Engineering zu einem zentralen Organisationsprinzip agentischer Softwareproduktion. KI-Agenten planen Aktionen, führen sie aus, beobachten deren Ergebnisse, bewerten diese Ergebnisse und bestimmen auf dieser Grundlage weitere Aufgaben~\cite{hassan2026agentic}. Die Entwicklung wird damit zu einem iterativen Prozess, in dem Ergebnisse einer Entwicklungsaktivität wieder zum Ausgangspunkt weiterer Entwicklungsaktivitäten werden. KI-Agenten können dabei zunehmend selbst SE Assets erzeugen und verändern; sie werden damit, wie in Abschnitt~\ref{de:sec:Selbsttransformation} beschrieben, zunehmend selbst zu Produzenten von SE Assets. 

Daten sowie Logs und Traces erhalten in diesem Prozess eine zusätzliche Bedeutung. Sie dokumentieren nicht mehr ausschließlich vergangene Systemzustände oder unterstützen die Fehlerdiagnose, sondern können als Kontext-, Erfahrungs- und Evidenzressourcen in nachfolgende Projektentscheidungen eingehen. Policies und Engineering Guidelines wiederum werden als maschineninterpretierbare Vorgaben für die Ausführung und Begrenzung agentischer Engineering-Prozesse dienen. Sie definieren damit nicht nur Randbedingungen an das entstehende System, sondern zunehmend auch Randbedingungen für dessen Entwicklung.

Damit verändert agentisches SE nicht lediglich einzelne Werkzeuge des SE. Es verändert die Beziehung zwischen den Aktivitäten des Entwicklungsprozesses und den dabei entstehenden SE Assets. Was im klassischen SE überwiegend Ergebnis einer Entwicklungsaktivität ist, kann im agentischen SE zugleich Ausgangspunkt und Steuerungsgrundlage für weitere Entwicklungsaktivitäten werden. Die SE Assets bilden damit zunehmend die Grundlage der rekursiven Softwareentwicklungsprozesse.

Diese Weiterentwicklungen in der Softwareproduktion erklären die besondere Bedeutung des agentischen SE für die vierte Digitalisierungsphase. Die entscheidende Veränderung besteht nicht darin, dass Maschinen erstmals Software erzeugen. Automatisierte Codegenerierung, modellbasierte Entwicklung und andere Formen der SE-Automatisierung existieren bereits seit Jahrzehnten. Neu ist vielmehr die zunehmende Verbindung unterschiedlicher SE-Aktivitäten in agentischen, feedbackbasierten Prozessen. Software kann dadurch zunehmend nicht nur automatisiert erzeugt, sondern innerhalb eines zusammenhängenden Prozesses analysiert, geprüft, verändert und weiterentwickelt werden.

\subsection{Von der manuellen Codeproduktion zur Gestaltung komplexer softwarebasierter soziotechnischer Systeme}

Mit der zunehmenden Automatisierung der Codeerzeugung verschiebt sich daher der Schwerpunkt der ingenieurmäßigen Wertschöpfung im SE. Es geht weniger um die manuelle Erstellung einzelner Codezeilen und zunehmend um die Präzisierung von Systemintentionen, die Architektur komplexer softwarebasierter Systeme, die Definition von Qualitätszielen, die V\&V sowie die Gestaltung und Governance von Mensch-KI- und Multi-Agenten-Produktionsprozessen.

Eine aufschlussreiche historische Parallele bietet die Entwicklung des Typographen bzw. Schriftsetzers~\cite{drucker1994visible}. Klassische Schriftsetzer beherrschten ein spezialisiertes Handwerk: Sie überführten Texte mithilfe von Bleilettern und später im Fotosatz in eine druckfähige typografische Form. Mit der Einführung des Desktop Publishing in den 1980er-Jahren wurden wesentliche Teile dieser Produktionsleistung digitalisiert und zugleich für andere Akteure zugänglich. Typographische Gestaltung konnte nun unmittelbar am Rechner erfolgen; die technische Trennung zwischen Autoren, Gestaltern, Setzern und der Druckvorstufe wurde teilweise aufgehoben. Damit verschwand nicht die Notwendigkeit von Typographie oder Gestaltung, wohl aber ein erheblicher Teil der spezialisierten manuellen Produktionsarbeit des Schriftsetzers. 
Die historische Entwicklung lässt sich dabei weniger als Verschwinden eines gesamten Berufs denn als Verschiebung seiner Wertschöpfungsposition verstehen. Maschinen und Softwarewerkzeuge übernahmen standardisierbare Produktionsschritte; menschliche Arbeit verlagerte sich auf Gestaltung, Auswahl, Komposition und Qualitätsbeurteilung. Digitalisierung beseitigt damit nicht notwendig eine Aufgabe als Ganzes, sondern kann die bisherige Trennung zwischen spezialisierten Produktionstätigkeiten und übergeordneter Gestaltung verändern.

Für das SE zeichnet sich eine strukturell ähnliche Entwicklung ab. Generative KI, insbesondere Large Language Models (LLMs), kann bereits heute aus natürlichsprachlich formulierten Anforderungen Code erzeugen; empirische Untersuchungen zeigen zugleich, dass die Qualität solcher Erzeugnisse nicht automatisch den Anforderungen professionell entwickelter Software entspricht (siehe auch Abschnitt~\ref{de:sec:Dynamiken}). Damit zeichnet sich eine mögliche Verschiebung vom codezentrierten zum intentzentrierten SE ab: Code wird vom primären Entwicklungsprodukt zu einem Element eines rekursiven Engineering-Prozesses.

Die Analogie zum Typographen macht zugleich deutlich, warum die Frage nach einer möglichen Ersetzung von Programmierern durch KI zu kurz greift. Entscheidend ist vielmehr, welche Tätigkeiten des SE automatisiert werden, welche neuen Tätigkeiten entstehen und welche Kompetenzen auf eine höhere Abstraktionsebene gehoben werden. So wie Desktop Publishing das Setzen von Text weitgehend automatisierte, ohne die Notwendigkeit von Typographie, Gestaltung und visueller Kommunikation aufzuheben, kann generative und agentische KI die Erstellung von Code weitgehend automatisieren, ohne die Notwendigkeit von SE aufzuheben.

Der Unterschied ist allerdings fundamental: Software ist nicht lediglich eine Form der Darstellung oder Verarbeitung von Information, sondern bildet die technische Grundlage ausführbarer und zunehmend gesellschaftlich kritischer soziotechnischer Systeme. Fehlerhafte oder unsichere Software kann wirtschaftliche Schäden verursachen, Menschen gefährden oder kritische Infrastrukturen beeinträchtigen. Mit der zunehmenden Automatisierung der Codeerzeugung verschwindet daher nicht die Verantwortung für Softwarequalität. Sie verlagert sich vielmehr zunehmend auf die Bestimmung von Zielen und Qualitätsanforderungen, die Überwachung und Bewertung von Entwicklungsergebnissen sowie die Governance teilautomatisierter Entwicklungsprozesse.

Gerade darin liegt die strategische Bedeutung des SE in der vierten Digitalisierungsphase. Wenn generative und agentische SE-Werkzeuge zunehmend Code und andere SE Assets erzeugen, wird SE nicht weniger relevant. Vielmehr wird es zur Disziplin, die diese teilautomatisierte Softwareproduktion orientiert, strukturiert, begrenzt, überprüft und verantwortbar macht. Die ingenieurmäßige Wertschöpfung verschiebt sich damit von der unmittelbaren Produktion von Code hin zur Gestaltung von Intention, Architektur, Qualität und Evolution komplexer softwarebasierter soziotechnischer Systeme.

Diese Verschiebung betrifft jedoch nicht allein Methoden und Werkzeuge des SE. Sie beruht auch auf einer Veränderung der Ressourcen und Bedingungen, unter denen Software produziert wird.

\subsection{Kognitive Infrastrukturen als Produktionsfaktor}
\label{de:subsec:Infrastruktur}

Über Jahrzehnte war kognitive Arbeit in der Digitalisierung an Personen, Teams und Organisationen gebunden. Zusätzliche Kapazitäten erforderten daher vor allem Ausbildung, Rekrutierung, Zusammenarbeit und den Aufbau von Wissen.

Generative und agentische KI verändern diese Produktionsbedingungen. Bestimmte kognitive Leistungen können zunehmend digital bereitgestellt, abgerufen und skaliert sowie unmittelbar in Arbeitsprozesse integriert werden. Ähnlich wie Cloud Computing Rechenleistung von lokaler Hardware entkoppelt hat~\cite{armbrust2009above}, ermöglicht KI die Bereitstellung bestimmter kognitiver Leistungen unabhängig von der unmittelbaren Verfügbarkeit einzelner Personen. Kognitive Kapazität wird damit in bestimmten Anwendungsbereichen zu einer elastisch einsetzbaren Ressource.

Diese technische Elastizität bildet die Grundlage für eine neue Form digitaler Infrastruktur. \textbf{Kognitive Infrastruktur} bezeichnet hier technische, organisatorische und institutionelle Ressourcen, durch die kognitive Leistungen wie Wissensverarbeitung, Generierung, Planung und Problemlösung als wiederverwendbare und skalierbare Leistungen bereitgestellt und in digitale Arbeitsprozesse integriert werden können. Das Konzept schließt an die Vorstellung einer kognitiven Infrastruktur als Zusammenspiel von Technologien, Diensten, Institutionen und Produkten an, die funktionale Elemente von Kognition bereitstellen~\cite{chester2023infrastructure}.

Als Produktionsfaktor ermöglicht eine solche Infrastruktur, kognitive Kapazität unabhängig von der unmittelbaren Verfügbarkeit einzelner Personen oder Teams in unterschiedlichen Größenordnungen in Produktions- und Entscheidungsprozesse einzubinden. Ihre produktive Wirkung hängt dabei nicht allein von der Leistungsfähigkeit der zugrunde liegenden KI ab, sondern ebenso von ihrer Verfügbarkeit, Skalierbarkeit, Zuverlässigkeit und Einbettung in organisatorische Prozesse. Damit schafft die kognitive Infrastruktur die Voraussetzung dafür, dass bestimmte kognitive Produktionsleistungen mit deutlich veränderten Grenzkosten bereitgestellt werden können. Die daraus resultierenden ökonomischen Konsequenzen für die Softwareproduktion werden im folgenden Abschnitt betrachtet.

\section{Die Ökonomie der kognitiven Grenzkosten}
\label{de:sec:Grenzkosten}

Die Digitalisierungsphasen lassen sich nicht nur danach unterscheiden, was und womit digitalisiert wird, sondern auch danach, welche Kosten digitaler Leistungen durch technologische Fortschritte sinken (siehe Tabelle~\ref{de:tab:Phasen}). Zunächst sanken die Kosten der Speicherung und Verarbeitung von Informationen, anschließend die Kosten der automatisierten Ausführung und Vernetzung digitaler Prozesse. Mit der digitalen Transformation wurden insbesondere Kommunikation, Koordination und datenbasierte Vorhersage kostengünstiger.

Die vierte Digitalisierungsphase setzt an einer weiteren Produktionsstufe an: Agentische KI senkt zunehmend die Grenzkosten bestimmter kognitiver Leistungen, die bislang überwiegend durch menschliche Arbeit erbracht wurden. Damit verändert sich auch die Kostenstruktur der Softwareproduktion. Neben die traditionellen Aufwendungen für Personal, Werkzeuge und Infrastruktur treten laufende Kosten für KI-Modellinferenz und agentische Interaktionen. Diese entstehen nicht nur bei der Codegenerierung, sondern auch bei Planung, Kontextverarbeitung, Werkzeugnutzung, Review, Test und Korrektur.

Die Bedeutung dieser variablen Kosten zeigt sich insbesondere bei agentischem SE: Eine aktuelle empirische Untersuchung von Multi-Agenten-Systemen im Software Development Life Cycle (SDLC) weist beispielsweise für die untersuchte Umgebung einen durchschnittlichen Anteil von 59,4~\% des gesamten Tokenverbrauchs für die iterative Code-Review-Phase aus; 53,9~\% des Verbrauchs entfielen auf Input-Tokens~\cite{salim2026tokenomics}. Die Kosten agentischer Softwareproduktion entstehen damit nicht allein bei der Erzeugung von Code, sondern über den gesamten iterativen Prozess der Kontextverarbeitung, Interaktion und Evaluation.

Damit verschiebt sich die ökonomische Betrachtung der Softwareproduktion. Entscheidend ist nicht allein, wie stark die Produktivität der Softwareentwicklung durch KI steigt, sondern wie sich die sinkenden Grenzkosten kognitiver Leistungen auf die laufenden Kosten, die Nachfrage und die erforderlichen Produktionsressourcen auswirken.

\subsection{Die Senkung kognitiver Grenzkosten}
\label{de:subsec:Senkung}

Die erste Digitalisierungsphase reduzierte zunächst die Kosten der Speicherung, Vervielfältigung und Verarbeitung digitaler Informationen. Die Prozess-Digitalisierung senkte anschließend die Kosten der automatisierten Ausführung und Reproduktion standardisierter digitaler Prozesse. Mit der digitalen Transformation wurden durch Vernetzung, Plattformen und datenbasierte Verfahren die Kosten der Koordination und insbesondere der datenbasierten Vorhersage erheblich reduziert. Agrawal u.a. interpretieren den ökonomischen Kern der KI entsprechend als eine wesentliche Senkung der Kosten von Vorhersagen~\cite{agrawal2018prediction}.

Die aktuelle Phase der digitalen Selbsttransformation erweitert diese Entwicklung um eine weitere Dimension: Agentische KI senkt zunehmend die Grenzkosten \textbf{kognitiver Synthese}. Unter kognitiver Synthese wird hier die Fähigkeit verstanden, aus Zielen, Wissen und Kontext neue SE Assets, Empfehlungen, Entscheidungen oder Handlungspläne zu erzeugen und ihre Eignung anhand von Rückmeldungen zu bewerten. Im Kontext der Softwareproduktion umfasst dies beispielsweise die Ableitung von Architekturentwürfen, Code, Tests oder Projektplänen aus Anforderungen und bestehendem Systemwissen.

Die Senkung digitaler Grenzkosten hat eine lange technische Vorgeschichte. Bereits 1965 beschrieb Moore   die langfristige Zunahme der Integrationsdichte integrierter Schaltkreise und revidierte seine Prognose 1975~\cite{moore1965cramming,moore1975progress}. Entscheidend ist dabei weniger die konkrete Verdopplungsrate als die über Jahrzehnte wesentlich sinkenden Kosten digitaler Rechenleistung. Fortschritte in Halbleitertechnologie, Rechnerarchitektur, Software und Algorithmen machten digitale Verarbeitung kontinuierlich leistungsfähiger und kostengünstiger~\cite{nordhaus2007two}. Mit generativer und agentischer KI setzt sich diese Entwicklung auf einer höheren Abstraktionsebene fort: Nicht nur Rechenoperationen, sondern zunehmend kognitive Leistungen werden digital ausführbar und skalierbar.

\subsection{Die variablen Kosten agentischer KI in der Softwareproduktion}

Für die folgende Betrachtung (siehe Tabelle~\ref{de:tab:Kosten}) werden zwei Dimensionen von Software unterschieden: die Art der Softwareerzeugung und die Art der Softwareausführung. Als \textbf{klassisch erzeugt} wird Software bezeichnet, deren Erstellung und Weiterentwicklung primär durch menschliche Softwareentwicklung erfolgt, auch wenn dabei automatisierte Entwicklungswerkzeuge oder KI-gestützte Assistenzsysteme eingesetzt werden. Als \textbf{agentisch erzeugt} wird Software bezeichnet, wenn KI-Agenten wesentliche Teile ihrer Erstellung, Prüfung oder Weiterentwicklung eigenständig und in mehrstufigen Arbeitsprozessen übernehmen. 
Davon unabhängig ist zu betrachten, wie die erzeugte Software ausgeführt wird. Klassische Software verarbeitet Eingaben auf Grundlage explizit implementierter Programmlogik und folgt damit einer zur Laufzeit weitgehend festgelegten Ausführungslogik. Agentische Software nutzt demgegenüber KI-Modelle zur eigenständigen Planung und Bearbeitung von Aufgaben und kann hierzu mehrere Inferenz-, Werkzeug- und Rückkopplungsschritte ausführen. „Klassisch“ und „agentisch“ bezeichnen damit unterschiedliche Ausprägungen der Erzeugungs- und Ausführungsweise. Aus der Kombination beider Dimensionen ergeben sich vier idealtypische Formen von Software: klassisch erzeugte klassische Software (kurz klassische Software), klassisch erzeugte agentische Software, agentisch erzeugte klassische Software und agentisch erzeugte agentische Software (kurz  agentische Software).

Sinkende Grenzkosten kognitiver Leistungen für die Softwareproduktion bedeuten dabei nicht, dass deren Bereitstellung kostenfrei wird. Vielmehr verändert sich die Kostenstruktur: Während bei klassischer Software die wesentlichen Entwicklungsaufwendungen weitgehend unabhängig von der Zahl ihrer Kopien anfallen und die Vervielfältigung mit sehr geringen Grenzkosten möglich ist, entstehen bei agentischer Software mit jeder zusätzlichen Inferenz variable Kosten für Rechenleistung, Energie und technische Infrastruktur (siehe Tabelle~\ref{de:tab:Kosten}). Dieser Zusammenhang wird dadurch verstärkt, dass in der Software(-produktion) typischerweise komplexe Aufgaben nicht durch einen einzelnen KI-Modellaufruf bearbeitet werden, sondern durch Sequenzen von Inferenzschritten, Werkzeugaufrufen sowie Bewertungs- und Rückkopplungsschleifen. Die Kosten einer kognitiven Leistung hängen damit nicht nur vom zugrunde liegenden KI-Modell, sondern auch von der Zahl und Komplexität der für ihre Erbringung erforderlichen Verarbeitungsschritte ab. Für die agentische Softwareproduktion bedeutet dies, dass ihre Ökonomie neben den Kosten der Entwicklung zunehmend auch durch die laufenden Kosten der agentischen Ausführung bestimmt wird.

Gleichzeitig sinken die Kosten dieser Ausführung erheblich. Die Höhe der Inferenzkosten hängt unter anderem von der Komplexität der Aufgabe, der Zahl und Art der KI-Modellaufrufe sowie von der eingesetzten Hard- und Softwareinfrastruktur ab. Untersuchungen von Epoch AI zeigen, dass die Inferenzpreise für ein gegebenes Leistungsniveau je nach Aufgabe innerhalb eines Jahres um Faktoren zwischen etwa neun und mehreren hundert sinken können~\cite{cottier2025llm}. Neuere Analysen führen diese Entwicklung sowohl auf Verbesserungen der Hardware- und Systemeffizienz als auch auf algorithmische Fortschritte und zunehmenden Wettbewerb zurück~\cite{gundlach2025price}. Damit entsteht eine für die vierte Digitalisierungsphase charakteristische ökonomische Dynamik: Kognitive Leistungen bleiben zwar mit positiven variablen Kosten verbunden, ihre Grenzkosten sinken jedoch rasch. Kognitive Verarbeitung kann dadurch in immer größerem Umfang automatisiert, wiederholt und skaliert werden. Entscheidend ist somit nicht allein die Höhe der Kosten einer einzelnen KI-Leistung, sondern das Verhältnis zwischen sinkenden Kosten je Leistungseinheit und der dadurch möglichen Ausweitung ihres wirtschaftlich sinnvollen Einsatzes.

Diese Entwicklung verändert zugleich die klassische Softwareökonomie (siehe Tabelle~\ref{de:tab:Kosten}). Bei klassischer Software fallen die wesentlichen Entwicklungsaufwendungen typischerweise vor der Nutzung an, während die Ausführung und Vervielfältigung eines bereits entwickelten Programms mit geringen Grenzkosten möglich ist. Bei agentisch erzeugter Software wird dagegen die Produktion selbst zu einem wiederkehrenden, rechenintensiven Inferenzprozess: Die Erstellung, Prüfung und Weiterentwicklung von SE Assets kann durch KI-Agenten erfolgen und verursacht damit für weitere Produktions- und Änderungsschritte variable Kosten. Sinkende Inferenzkosten machen diese Form der Produktion zunehmend wirtschaftlich attraktiv, beseitigen jedoch nicht die damit verbundenen Folgekosten.

Sinkende Grenzkosten der Codegenerierung können zugleich einen gegenläufigen Kosteneffekt auslösen: Je günstiger SE Assets erzeugt werden können, desto größer kann die Menge des erzeugten und damit zu prüfenden, zu integrierenden und zu wartenden Codes werden. Eine großskalige Untersuchung von 304.362 verifizierten KI-generierten Commits aus 6.275 GitHub-Repositories identifizierte 484.606 durch diese Commits eingeführte Probleme; 89,1~\% davon entfielen auf Code Smells. Bei mehr als 15~\% der untersuchten Commits der betrachteten KI-Werkzeuge wurde mindestens ein solches Problem eingeführt. Besonders relevant ist, dass 24,2~\% der nachverfolgten, in den KI-generierten Commits eingeführten Probleme in der jeweils letzten untersuchten Revision weiterhin bestanden~\cite{liu2026debt}. Sinkende Kosten der Codegenerierung führen damit nicht automatisch zu sinkenden Gesamtkosten der Softwareproduktion. Sie können vielmehr die Kostenstruktur innerhalb des Softwarelebenszyklus verschieben: Einsparungen bei der Codegenerierung können mit zusätzlichen Aufwendungen für V\&V, Integration, Fehlerbehebung und langfristige Wartung einhergehen. Entscheidend für die ökonomische Wirkung agentischer Softwareproduktion ist damit nicht allein, wie billig Code erzeugt werden kann, sondern wie sich die Gesamtkosten über seinen Lebenszyklus entwickeln.

\begin{table}
    \centering
    \renewcommand{\arraystretch}{1.2}
    \setlength{\tabcolsep}{2pt}
    \begin{tabular}{|>{\raggedright\arraybackslash}p{2.2cm}|>{\raggedright\arraybackslash}p{2.2cm}|>{\raggedright\arraybackslash}p{2.2cm}|>{\raggedright\arraybackslash}p{2.2cm}|>{\raggedright\arraybackslash}p{2.2cm}|}
    \hline 
        \textbf{Phase} & 
        \textbf{Klassische Software} & 
        \textbf{Agentisch erzeugte klassische Software} & 
        \textbf{Klassisch erzeugte agentische Software} & 
        \textbf{Agentische Software} \\
    \hline 
        Entwicklung und Weiterentwicklung & 
        überwiegend fixe Gesamtkosten & 
        variable KI-Inferenzkosten & 
        überwiegend fixe Gesamtkosten & 
        variable KI-Inferenzkosten \\
    \hline 
        Ausführung und Nutzung & 
        geringe Grenzkosten & 
        geringe Grenzkosten & 
        variable KI-Inferenzkosten & 
        variable KI-Inferenzkosten \\
    \hline 
    \end{tabular}
    \caption{Vereinfachte Kostenmerkmale verschiedener Formen von Software}
    \label{de:tab:Kosten}
\end{table}

\subsection{Eine mögliche Expansion der Softwareproduktion}

Des Weiteren führen sinkende Grenzkosten der Softwareproduktion nicht notwendig zu einer Verringerung der Gesamtnachfrage nach Software. Vielmehr können Produktivitätssteigerungen einen Rebound-Effekt~\cite{jevons1934william} auslösen, indem sie Anwendungen wirtschaftlich realisierbar machen, deren Entwicklung bei höheren Kosten unterblieben wäre. In diesem Sinne lässt sich das Jevons-Paradoxon als heuristischer Referenzrahmen auf die Softwareproduktion übertragen: Eine effizientere Produktion kann die Nutzung eines Gutes ausweiten, anstatt sie zu reduzieren~\cite{sorrell2009jevons}.

Für Software ist dabei insbesondere die Erweiterung des adressierbaren Anwendungsraums relevant. Sinkende Entwicklungskosten machen nicht nur zusätzliche Anwendungen in bestehenden Märkten wirtschaftlich, sondern ermöglichen auch Lösungen für individuelle, situative oder zeitlich begrenzte Anforderungen, deren Entwicklungsaufwand bislang den erwarteten Nutzen überstieg. Die ökonomisch relevante Grenze verschiebt sich damit von der Frage, ob eine Softwarelösung technisch realisierbar ist, zu der Frage, für welche zusätzlichen Zwecke ihre Entwicklung wirtschaftlich sinnvoll wird.

Agentisches SE kann diese Entwicklung verstärken, indem es Software nicht nur effizienter erzeugt, sondern auch deren Anpassung an wechselnde Kontexte unterstützt. Software kann dadurch von einem vorab entwickelten und anschließend weitgehend statischen Artefakt zu einer anpassungsfähigen Ressource werden. Erste empirische Untersuchungen zeigen bereits substanzielle Produktivitätsgewinne durch generative KI bei bestimmten wissensintensiven Tätigkeiten~\cite{brynjolfsson2025generative}. Ob die dadurch sinkenden Produktionskosten langfristig zu einer überproportionalen Ausweitung der Gesamtnachfrage nach Software führen, ist jedoch eine empirische Frage und nicht durch das Produktivitätsargument allein bestimmt.

\subsection{Die Verschiebung ökonomischer Knappheiten}

Eine Ausweitung der Softwareproduktion verändert zugleich die relative Bedeutung der für die Softwareproduktion benötigten Ressourcen und verändert die Struktur der Knappheiten innerhalb der Softwareproduktion. Wenn die Erzeugung von Software zunehmend automatisiert werden kann, verliert die reine Produktionskapazität als Differenzierungsfaktor an Bedeutung~\cite{zhang2026economics}. Zugleich können Ressourcen und Fähigkeiten, die für die Auswahl, Bewertung, Absicherung und Einbettung erzeugter Systeme erforderlich sind, relativ knapper werden.

Hierzu zählen insbesondere verlässliche Daten, Rechen- und Energieinfrastruktur sowie Kompetenzen in V\&V, Qualitätssicherung, Sicherheit, Datenschutz und Governance. Hinzu kommt menschliche Urteilskraft bei der Spezifikation von Zielen, der Bewertung von Ergebnissen und der Entscheidung über den Einsatz von Systemen in konkreten sozialen und institutionellen Kontexten.

Die ökonomische Entwicklung der Softwareproduktion ist daher nicht als Übergang von Knappheit zu allgemeiner Abundanz zu verstehen. Vielmehr verschiebt sich die relative Knappheit zwischen unterschiedlichen Produktionsfaktoren: Während die Kosten der Erzeugung von Software sinken und deren Angebot ausgeweitet werden kann, können komplementäre Ressourcen und Fähigkeiten zu relativen Engpässen werden. Die zunehmende Verfügbarkeit von Code führt damit nicht zu einer Aufhebung ökonomischer Knappheit, sondern zu ihrer partiellen Verlagerung innerhalb der Softwareproduktion.

\section{Soziotechnische Dynamiken der Softwareproduktion}
\label{de:sec:Dynamiken}

Mit agentischer KI verändert sich ebenso die soziotechnische Organisation der Softwareproduktion: Arbeitsteilung, Organisationsstrukturen und Kontrollmechanismen müssen neu gestaltet werden. Dabei stellt sich zugleich die Frage, worin die ingenieurwissenschaftliche Leistung in der Softwareentwicklung künftig besteht.

\begin{figure}
    \centering
    \includegraphics[width=\linewidth]{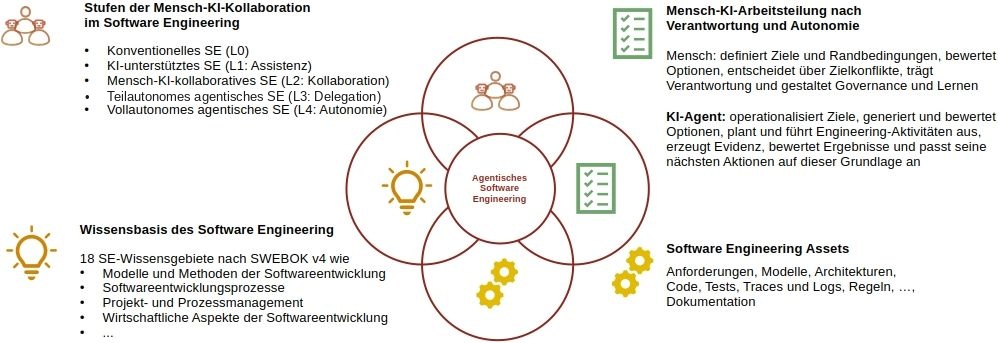}
    \caption{Agentisches SE als soziotechnisches System}
    \label{de:fig:AgentischesSE}
\end{figure}

Während frühere KI-basierte Assistenzsysteme im SE vor allem einzelne Tätigkeiten wie die Erzeugung oder Transformation von SE-Artefakten unterstützten, können agentische SE-Systeme zunehmend zusammenhängende Entwicklungsaufgaben über mehrere Schritte hinweg ausführen. Sie analysieren Anforderungen und Code-Repositories, erstellen Pläne, verändern Dateien, nutzen Softwarewerkzeuge, führen Tests aus und passen ihr Vorgehen auf Grundlage der Ergebnisse rekursiv an. Systeme wie SWE-agent demonstrieren diese Verschiebung von der Code-Assistenz zur repositoryweiten Bearbeitung von Software-Engineering-Aufgaben~\cite{yang2024sweagent}.

Damit verändert sich die Arbeitsteilung zwischen Mensch und Maschine: Statt einzelne Aktivitäten auszuführen, delegiert der Mensch zunehmend zusammenhängende Aufgaben an ein agentisches SE-System und steuert dessen Handlungsspielraum über Ziele, Kontext, Randbedingungen und Kontrollmechanismen. Die Entwicklung verschiebt sich damit von der punktuellen Assistenz über die Zusammenarbeit und Delegation hin zu zunehmend autonomen Entwicklungsaktivitäten (siehe Abbildung~\ref{de:fig:AgentischesSE}).

\subsection{Von der Programmierung zum Supervisory Engineering}

Empirische Untersuchungen zur Nutzung von KI-Coding-Assistenten stützen die These der Verschiebung zu zunehmend autonomen Entwicklungsaktivitäten: Eine longitudinale Studie unter professionellen Softwareentwicklern und Softwareentwicklerinnen zeigt, dass 82~\% der Befragten weniger Zeit mit dem Schreiben von Code verbringen und sich die Tätigkeit insgesamt von der Erstellung hin zu V\&V-Aktivitäten verschiebt~\cite{vella2026impact}. Die Autoren führen hierfür den Begriff \textbf{Supervisory Engineering Work} als eine Form ingenieurwissenschaftlicher Arbeit, die die Steuerung, Evaluation und Korrektur von KI-generierten SE Assets umfasst, ein. 

Dabei werden Softwareingenieure nicht zu bloßen Managern von KI-Agenten, sondern übernehmen zunehmend die Verantwortung für die Gestaltung und Überwachung des Produktionsprozesses, in dem agentische KI-Systeme (Teil-)Aufgaben (teil-)autonom ausführen. Supervisory Engineering umfasst insbesondere die Definition von Arbeitsaufträgen und Randbedingungen, die Steuerung von Agenten-Workflows, die Evaluation ihrer Ergebnisse sowie die Identifikation und Korrektur von Fehlern, die auf der Ebene einzelner Code-Komponenten nicht unmittelbar erkennbar sind~\cite{vella2026impact}.

Damit verschiebt sich zudem die zeitliche und kognitive Struktur der Entwicklungsarbeit. Die unmittelbare Produktion von Code nimmt ab, während Review, V\&V, Fehleranalyse und die wiederholte Interaktion mit KI-Agenten an Bedeutung gewinnen. Vella et al.~\cite{vella2026impact}~zeigen dabei ein bemerkenswertes Produktivitäts-Erfahrungs-Paradoxon: 84~\% der Befragten berichteten sowohl zu Beginn als auch sechs Monate später von Produktivitätsverbesserungen, während sich gleichzeitig der Anteil der Personen, die eine Verschlechterung mindestens einer Dimension der Developer Experience wahrnahmen, von 14~\% auf 27~\% erhöhte. Insbesondere Flow und wahrgenommene kognitive Belastung verschlechterten sich, während sich Feedbackschleifen verbesserten. 

Des Weiteren relativieren die Ergebnisse in~\cite{vella2026impact} eine rein outputorientierte Betrachtung von Produktivität. Eine höhere Geschwindigkeit bei der Erzeugung von Software bedeutet nicht notwendigerweise eine bessere Arbeitserfahrung oder eine höhere Effektivität des gesamten Entwicklungsprozesses. In einem anderen empirischen Setting zeigte eine randomisierte kontrollierte Studie sogar eine um 19~\% längere Bearbeitungszeit bei erlaubter Nutzung damaliger KI-Coding-Werkzeuge~\cite{becker2025measuring}. Zugleich schätzten die Entwickler ihre eigene Produktivitätssteigerung nach Abschluss der Studie auf durchschnittlich 20~\%. Die Autoren betonen allerdings, dass diese Ergebnisse als kontextabhängige Momentaufnahme zu verstehen sind und nicht auf die gesamte Softwareentwicklung generalisiert werden können~\cite{becker2025measuring}.

Mit dem Wandel der Softwareproduktion verändert sich auch das Kompetenzprofil der Softwareingenieure. Entscheidend werden zunehmend Systemverständnis, Anforderungspräzision, Architekturkompetenz, V\&V-Kompetenz und die Fähigkeit, agentische KI-Systeme in einem gegebenen technischen und organisatorischen Kontext wirksam zu orchestrieren.

So besteht die zentrale Veränderung vom klassischen SE zum agentischen SE weniger in einer einfachen Substitution menschlicher Entwicklungsarbeit als in einer Verschiebung von Programmierung zu Konzeption und Supervision: Die Ingenieure gestalten zunehmend die Bedingungen, unter denen agentische KI-Systeme Software programmieren, und tragen zugleich die Verantwortung für die technische Angemessenheit und die Qualität des Ergebnisses.

\subsection{Die Rekonfiguration von Teams und Organisationsstrukturen}

Wenn agentische KI-Systeme einen größeren Anteil der Entwicklungsaktivitäten übernehmen, verändert sich zwangsläufig auch die Organisation der Softwareproduktion. Aufgaben, die bislang auf mehrere spezialisierte Rollen verteilt waren, können zunehmend von kleineren menschlichen Einheiten koordiniert werden, die auf eine Vielzahl spezialisierter KI-Agenten zurückgreifen. Dadurch kann sich der Handlungsspielraum einzelner Ingenieure erhöhen.

Aus dieser Entwicklung lässt sich jedoch noch keine empirisch abgesicherte neue Standardform der Teamorganisation ableiten. Die häufig formulierte Vorstellung hochintegrierter Kleinstteams oder sogenannter Super-Zellen, die mithilfe von (Schwärmen von) KI-Agenten große Teile des Software-Stacks bearbeiten, ist daher zunächst als Organisationshypothese zu verstehen. Ihre Realisierung hängt unter anderem von der Zuverlässigkeit der KI-Agenten und ihrer Einbettung in die Softwareproduktion, der Komplexität der zu entwickelnden softwarebasierten Systeme, der erforderlichen Branchenexpertise und dem regulatorischen und sicherheitsrelevanten Kontext ab.

Wahrscheinlicher als eine generelle Auflösung von Spezialisierungen der SE-Rollen und -Teams erscheint zunächst eine Verschiebung ihrer Funktionen. Plattform- und Infrastrukturteams könnten an Bedeutung gewinnen, weil sie nicht mehr nur Entwicklungs- und Laufzeitinfrastrukturen bereitstellen, sondern zunehmend auch die technischen und organisatorischen Rahmenbedingungen für eine agentische Softwareproduktion schaffen: Kontextbereitstellung, Zugriffsrechte, Werkzeuge, Evaluationsinfrastrukturen, Monitoring, Sicherheitsmechanismen und Governance.
Damit verschiebt sich zudem die Bedeutung von IT-Plattformen: Sie werden von einer unterstützenden Infrastruktur zu einer möglichen organisatorischen Kontrollinstanz der agentischen Softwareproduktion.

\subsection{Von direkter Kontrolle zu systemischer Governance}

Je autonomer KI-Agenten handeln, desto weniger kann Governance über die unmittelbare menschliche Kontrolle einzelner Entwicklungsschritte erfolgen. Menschliche Intention muss daher zunehmend vorab in persistenten und maschineninterpretierbaren SE Assets verankert werden. Die Steuerung verschiebt sich damit von der direkten Beaufsichtigung einzelner Arbeitsschritte hin zur Gestaltung von Regeln, Grenzen und überprüfbaren Zielvorgaben für das agentische SE.

Dadurch verschmilzt Governance unmittelbar mit der Softwareproduktion: Anstatt Regeln und Sicherheitsrichtlinien rein deklarativ zu dokumentieren, werden sie als maschineninterpretierbare SE Assets direkt in den Entwicklungs- und Ausführungsprozess integriert. KI-Agenten bewegen sich autonom innerhalb dieser vordefinierten Leitplanken, während ihre Ergebnisse wiederum die notwendige Evidenz für automatisierte Freigaben und fundierte menschliche Entscheidungen liefern.

Dabei ist zwischen technischer Governance und regulatorischer Compliance zu unterscheiden. Der EU AI Act etabliert für bestimmte KI-Systeme und deren Anbieter bzw. Betreiber regulatorische Anforderungen, bildet jedoch kein vollständiges Governance-Modell für agentisches SE. Für produktionsreife agentische Softwaresysteme ist vielmehr ein mehrschichtiges Kontrollregime erforderlich, das menschliche Intention, Anforderungen und Spezifikation mit KI-Agenten-Aktionen, erzeugten SE Assets, V\&V sowie dem Verhalten des Systems zur Laufzeit verknüpft. Traceability wird damit zu einem zentralen Mechanismus: Sie muss nachvollziehbar machen, wie Anforderungen und Gestaltungsentscheidungen in KI-Agenten-Aktionen und erzeugte SE Assets eingehen und wie deren Konformität und Qualität durch V\&V überprüft werden.

Mit zunehmender Autonomie verschiebt sich damit das Verhältnis von Produktions- und Kontrollkapazität. Agentische Systeme können Software in einer Geschwindigkeit und Menge erzeugen und verändern, die mit herkömmlichen Verfahren für V\&V möglicherweise nicht mehr Schritt hält. Entscheidend ist daher nicht allein die Menge des erzeugten Codes, sondern die Frage, ob die Kontrollkapazität mit der wachsenden Produktionskapazität Schritt halten kann~\cite{cotroneo2025human}. Wächst die Fähigkeit zur Erzeugung von Software schneller als die Fähigkeit, ihre Korrektheit, Sicherheit, Wartbarkeit und Systemverträglichkeit zu überprüfen, kann V\&V zum kritischen Engpass der Softwareproduktion werden~\cite{alenezi2026rethinking}.

Für die Sicherheit ist diese Entkopplung besonders relevant. Perry u.~a. zeigten in einer kontrollierten Nutzerstudie, dass Teilnehmerinnen und Teilnehmer mit einem KI-Code-Assistenten bei sicherheitsbezogenen Programmieraufgaben weniger sicheren Code produzierten und zugleich eher davon ausgingen, sicheren Code erzeugt zu haben~\cite{perry2023users}. Das Risiko entsteht damit nicht allein durch fehlerhafte oder unsichere KI-generierte SE Assets, sondern auch durch eine mögliche Überschätzung der eigenen Kontrollfähigkeit. Mit sinkenden Kosten und steigender Geschwindigkeit der Codegenerierung gewinnt daher die unabhängige und systematische V\&V an Bedeutung.

Die zentrale systemische Herausforderung der agentischen Softwareproduktion besteht somit in der Synchronisierung von Produktions- und Kontrollkapazität. Je stärker Softwareproduktion automatisiert und beschleunigt wird, desto leistungsfähiger müssen auch die Mechanismen ihrer V\&V und Governance werden. Erst wenn automatisierte Tests, statische und dynamische Analysen, Sicherheitsüberprüfungen, Laufzeitkontrollen und menschliche Urteilskraft mit der Geschwindigkeit agentischer Softwareproduktion Schritt halten können, lässt sich Code-Abundanz in nachhaltige Softwareproduktivität übersetzen. Agentische Softwareproduktion erfordert damit nicht weniger Kontrolle, sondern eine andere Form von Kontrolle: weniger unmittelbare Beaufsichtigung einzelner Arbeitsschritte und mehr explizite, maschinenverarbeitbare Regeln, kontinuierliche V\&V und nachvollziehbare Kontrollketten.

\section{Pfadabhängigkeiten und Gestaltungsoptionen der digitalen Selbsttransformation}
\label{de:sec:Gestaltung}

Die vierte Digitalisierungsphase ist keine technologisch determinierte Entwicklung. Zwar schaffen generative und agentische KI neue technische Möglichkeiten, welche dieser Möglichkeiten tatsächlich wirksam werden, welche Akteure von ihnen profitieren und welche Abhängigkeiten entstehen, wird jedoch durch bestehende technische (siehe Abschnitt~\ref{de:sec:Software}), ökonomische (siehe Abschnitt~\ref{de:sec:Grenzkosten}), organisationale und institutionelle  (siehe Abschnitt~\ref{de:sec:Dynamiken}) Strukturen mitbestimmt. 

Die zentrale Gestaltungsfrage lautet deshalb nicht allein, wie KI möglichst reibungslos adaptiert werden kann. Entscheidend ist vielmehr, wem die Fähigkeit zur souveränen Entwicklung, Veränderung und Kontrolle digitaler Systeme obliegt.  Die Gestaltungsaufgabe besteht darin, positive Entwicklungspfade zu ermöglichen, ohne neue technologische und institutionelle Lock-ins zu erzeugen. Für Deutschland und Europa bedeutet dies insbesondere, eigene Kompetenzen entlang der Softwareproduktionskette aufzubauen und die Fähigkeit zu erhalten, zwischen unterschiedlichen technologischen Pfaden wählen zu können.

\subsection{Die Software Engineering-Lotterie}

Technologische Entwicklung verläuft nicht allein als Auswahl jeweils überlegener Technologien. Sie wird durch Rückkopplungen zwischen Technologien, Infrastrukturen, Investitionen, Kompetenzen und Ökosystemen geprägt. Arthur beschreibt solche Entwicklungen als pfadabhängige Prozesse, in denen positive Rückkopplungen und selbstverstärkende Mechanismen dazu führen können, dass sich einmal eingeschlagene technologische Pfade stabilisieren, selbst wenn alternative Lösungen nicht grundsätzlich unterlegen sind~\cite{arthur1989competing}.

Hookers Konzepte der Hardware Lottery und Software Lottery verdeutlichen diese Dynamik für die KI-Forschung. Forschungsrichtungen setzen sich teilweise deshalb durch, weil sie besonders gut auf der jeweils verfügbaren Hardware und Software implementierbar sind. Technologische Möglichkeiten werden dadurch nicht unabhängig von ihrer materiellen Infrastruktur bewertet, sondern durch diese mitgeprägt~\cite{hooker2021hardware}. Der Aufstieg des Deep Learning illustriert diesen Zusammenhang: Die Kombination aus geeigneten Algorithmen, großen Datensätzen und GPU-basierter Rechenleistung ermöglichte Skalierungen, die zuvor nicht praktikabel waren~\cite{krizhevsky2012imagenet}. Damit wurde ein Entwicklungspfad begünstigt, in dem leistungsfähigere Hardware, größere Modelle, steigende Nachfrage und wachsende Softwareökosysteme einander zunehmend verstärkten.

Für die vierte Digitalisierungsphase bestehen so Pfadabhängigkeiten: Neben der Hardware wird zunehmend auch die Softwareproduktionsinfrastruktur selbst zu einem Faktor technologischer Pfadabhängigkeit~\cite{gururaja2023build}. KI-Modelle, KI-Agenten-Frameworks, Entwicklungsumgebungen, Werkzeuge, Trainingsdaten, Benchmarks, Engineering"=Praktiken und Kompetenzen verstärken sich gegenseitig. Wer beispielsweise über leistungsfähige KI-Coding-Agenten verfügt, aber nicht über geeignete SE Assets, Prüfmechanismen und organisatorische Kompetenzen, kann deren Potenzial nur begrenzt ausschöpfen. Umgekehrt erzeugt die breite Nutzung bestimmter KI-Agenten- und Plattformarchitekturen Anreize, Entwicklungsprozesse und Kompetenzen genau auf diese Systeme auszurichten.

Für die Softwareproduktion lässt sich diese Dynamik als \textbf{Software Engineering Lottery} beschreiben: Nicht nur die verfügbare Hardware und Software beeinflussen, welche KI-Systeme erfolgreich entwickelt werden können; auch die vorhandenen SE-Kompetenzen, SE Assets und Produktionsumgebungen prägen, welche Formen agentischer Softwareproduktion sich praktisch durchsetzen und skalieren lassen. Technologien werden damit nicht unabhängig von den Produktionsbedingungen bewertet und eingesetzt, sondern durch diese mitgeprägt. Die Fähigkeit, Software durch KI entwickeln zu lassen, wird selbst zu einer infrastrukturellen Kompetenz.

Für Europa folgt daraus eine zentrale Gestaltungsoption: Technologische Souveränität darf nicht bei Halbleitern und Rechenzentren enden. Sie muss auch die Softwareproduktionskette umfassen. Dazu gehören offene Entwicklungsumgebungen, interoperable Werkzeug-Schnittstellen, hochwertige Software- und Engineering-Daten, europäische Benchmarks, reproduzierbare Bewertungsverfahren sowie Forschung an alternativen Agenten- und Systemarchitekturen. Gerade in Bereichen, in denen Standards und Ökosysteme noch nicht vollständig konsolidiert sind, bestehen Ansatzpunkte, alternative Entwicklungspfade zu fördern und Abhängigkeiten frühzeitig zu begrenzen.

\subsection{Das Softwareproduktions-Lock-in}

Die ökonomische Dimension dieser Entwicklung ist eng mit der Konzentration der KI-Infrastrukturen verbunden. Frontier-KI erfordert hohe Investitionen in Rechenleistung, Modelle, Daten, Energie und hochqualifiziertes Personal. Skaleneffekte können dadurch zu einer Konzentration auf wenige Anbieter führen~\cite{korinek2025concentrating}. Gleichzeitig sinken die Kosten bestimmter Formen der Wissenssynthese und der Softwareproduktion. Diese gegenläufige Entwicklung – Konzentration der Produktionsmittel bei gleichzeitig sinkenden Grenzkosten einzelner Leistungen – kann eine asymmetrische Wertschöpfungsstruktur begünstigen, in der wenige Anbieter zentrale Produktionsressourcen kontrollieren, während deren Leistungen von einer großen Zahl von Nutzern und Organisationen eingesetzt werden.

Für die Softwareproduktion ist diese Asymmetrie besonders relevant. Wenn KI-Agenten zunehmend Software erzeugen, entsteht ökonomische Abhängigkeit nicht mehr nur durch die Nutzung fertiger Software, sondern zunehmend auch durch die Bindung der Softwareproduktion an bestimmte Technologien und Infrastrukturen. Unternehmen und öffentliche Organisationen können ihre Entwicklungsprozesse, Wissensbestände, Werkzeuge, Tests und Kompetenzen auf bestimmte KI-Modelle, KI-Agenten-Frameworks oder KI-Plattformen ausrichten. Mit zunehmender Integration können daraus Wechselkosten entstehen: Proprietäre Schnittstellen, spezifische Daten- und Kontextstrukturen, angepasste Entwicklungsprozesse, erworbene Kompetenzen und komplementäre Werkzeuge können den Wechsel zu alternativen Anbietern technisch aufwendig und ökonomisch unattraktiv machen. Forschung zu Cloud Computing zeigt, dass insbesondere mangelnde Interoperabilität und Portabilität sowie proprietäre Schnittstellen wesentliche Mechanismen von Vendor Lock-in darstellen~\cite{opara2016critical,kaur2017interoperability}.

Für die agentische Softwareproduktion lässt sich diese Form der Abhängigkeit als \textbf{Softwareproduktions-Lock-in} konzeptualisieren. Eine Organisation kann dabei auf unterschiedlichen Ebenen abhängig werden: durch die Bindung ihrer Softwaresysteme an bestimmte Plattformen (System-Lock-in), durch die Bindung ihrer technischen Infrastruktur an bestimmte Anbieter (Infrastruktur-Lock-in) oder durch die Bindung ihrer Softwareproduktionsprozesse an bestimmte KI-Modelle und Produktionsumgebungen (Softwareproduktions-Lock-in). Letzterer kann entstehen, wenn wesentliche Teile der eigenen Softwareproduktionsfähigkeit – etwa Entwicklungswissen, Agenten-Workflows, Evaluationsverfahren, Kontext- und Wissensbestände oder technische Schnittstellen – auf die Leistungen eines bestimmten KI-Modells oder einer bestimmten KI-Plattform zugeschnitten sind. Je höher die daraus entstehenden Wechselkosten und je geringer die Interoperabilität mit alternativen Produktionsumgebungen, desto stärker kann diese Abhängigkeit werden. 
Agentische KI kann damit Lock-in eine Ebene nach vorne verschieben: von der Abhängigkeit von digitalen Systemen und Infrastrukturen hin zur Abhängigkeit von den Bedingungen, unter denen digitale Systeme produziert werden.

Damit verschiebt sich auch die Frage der digitalen Souveränität von der Kontrolle über fertige Anwendungen hin zur Kontrolle über die Produktionsbedingungen digitaler Systeme. Souveränität bedeutet dabei nicht die vollständige Unabhängigkeit von externen Anbietern, sondern die Fähigkeit, zwischen relevanten Alternativen wählen, Abhängigkeiten bewerten und bei Bedarf Produktionsbedingungen verändern oder Anbieter wechseln zu können. Eine \textbf{souveräne Softwareproduktion} muss daher Wahlmöglichkeiten auf mehreren Ebenen erhalten: bei Recheninfrastrukturen, Daten, Softwareentwicklungsumgebungen, KI-Modellen, KI-Agenten-Frameworks, Schnittstellen und Evaluationsverfahren.

Zugleich darf die sinkende Produktionsbarriere nicht mit einer automatischen Steigerung gesamtwirtschaftlicher Produktivität gleichgesetzt werden. Generative und agentische KI weist wesentliche Merkmale einer General-Purpose Technology (GPT) auf: Ihre produktive Nutzung erfordert komplementäre Investitionen in Prozesse, Organisation, Geschäftsmodelle und Humankapital. Brynjolfsson, Rock und Syverson beschreiben diesen Zusammenhang als Productivity J-Curve. In der Einführungsphase können die Produktivitätseffekte einer GPT hinter ihren technischen Möglichkeiten zurückbleiben, weil komplementäre immaterielle Investitionen zunächst aufgebaut werden müssen~\cite{brynjolfsson2021productivity}. Die Übertragung dieses Mechanismus auf generative KI wird auch in jüngeren Analysen als relevante Erklärung für die verzögerte Realisierung von Produktivitätspotenzialen diskutiert~\cite{calvino2025generative}.

Für die digitale Selbsttransformation bedeutet dies: Der entscheidende Produktivitätshebel liegt nicht in der isolierten Einführung von KI-Coding-Agenten. Er entsteht vielmehr durch die Reorganisation der Softwareproduktion – von Anforderung und Spezifikation über Architektur und Implementierung bis zu Test, Betrieb, Wartung und Evolution. Agentisches SE ist damit nicht primär eine neue Werkzeugklasse, sondern ein neues Produktionsparadigma, dessen Produktivitätspotenzial erst durch komplementäre organisatorische, technische und kompetenzbezogene Veränderungen realisiert werden kann.

Empirische Ergebnisse stützen bereits die Annahme erheblicher Produktivitätspotenziale, zeigen aber zugleich deren Kontextabhängigkeit. Randomisierte Feldexperimente mit 4.867 Softwareentwicklern unter anderem bei Microsoft und Accenture fanden im Mittel eine Steigerung der Zahl abgeschlossener Aufgaben um 26,08~\% bei Nutzung eines KI-Coding-Assistenten; zugleich variierten die Ergebnisse zwischen den einzelnen Experimenten deutlich~\cite{cui2026effects}. Die Ergebnisse belegen damit einen Produktivitätseffekt unter den untersuchten Bedingungen, lassen sich jedoch nicht unmittelbar als proportionale Steigerung der Softwarequalität oder des gesamtwirtschaftlichen Ergebnisses interpretieren. Gerade deshalb gewinnt SE als Disziplin der Systemgestaltung, Qualitätssicherung und organisationalen Einbettung an Bedeutung: Je günstiger die Produktion von Software und ihrer SE Assets wird, desto wichtiger wird die Fähigkeit, deren Qualität, Integration, Betrieb und langfristige Weiterentwicklung systematisch zu gewährleisten.

\subsection{Digitale Veränderungsfähigkeit}

Aus den beschriebenen technologischen und ökonomischen Pfadabhängigkeiten folgt eine weitere Voraussetzung der digitalen Selbsttransformation: Organisationen müssen in der Lage sein, ihre digitalen Produktionsbedingungen kontinuierlich anzupassen und neu zu konfigurieren~\cite{li2019dynamic,teece2007explicating}. Dies stellt eine besondere Herausforderung dar, weil bestehende IT-Landschaften, Beschaffungsmodelle, Rollenbilder, Governance-Strukturen und Qualifikationsprofile selbst pfadabhängig sind. Je stärker diese Strukturen auf bestimmte Technologien und Produktionsweisen ausgerichtet sind, desto höher können die organisatorischen und ökonomischen Kosten ihrer Veränderung werden.

Die Digitalisierung wurde und wird vielfach projektförmig organisiert: Eine Anwendung wird eingeführt, ein Prozess digitalisiert oder eine Plattform aufgebaut. In der vierten Digitalisierungsphase reicht dieses Muster allein nicht mehr aus. Wenn Software zunehmend dynamisch erzeugt, geprüft und verändert werden kann, wird ihre kontinuierliche Evolution selbst zu einem Bestandteil der digitalen Produktionsweise. Damit verschiebt sich die organisationale Herausforderung von der erfolgreichen Durchführung einzelner Digitalisierungsprojekte hin zur dauerhaften Fähigkeit, digitale Systeme und ihre Produktionsbedingungen weiterzuentwickeln.

Dies betrifft insbesondere öffentliche Institutionen. Föderale Zuständigkeiten, heterogene IT-Landschaften, unterschiedliche Beschaffungslogiken und fragmentierte Verantwortlichkeiten können die kontinuierliche Weiterentwicklung digitaler Systeme erschweren. Der DigitalPakt Schule~\cite{digitalpaktschule2026bilanz} kann dabei als Beispiel für die Grenzen einer primär infrastrukturellen Perspektive dienen: Technische Ausstattung schafft notwendige Voraussetzungen, erzeugt aber für sich genommen noch keine dauerhafte Fähigkeit zur digitalen Veränderung. Auf die Softwareproduktion übertragen bedeutet dies: Der Besitz von KI-Modellen generiert erst dann Wertschöpfung, wenn komplementäre organisatorische Fähigkeiten und Governance-Strukturen etabliert werden.

Für die vierte Digitalisierungsphase verschärft sich diese Anforderung. Organisationen müssen nicht nur digitale Systeme betreiben, sondern zunehmend ihre eigene digitale Produktionsfähigkeit organisieren und weiterentwickeln. Dazu gehören die Fähigkeit zur Formulierung präziser Anforderungen und Spezifikationen, zur Architekturentwicklung, zum Einsatz und zur Orchestrierung von KI-Agenten, zur kontinuierlichen V\&V sowie zur Bewertung von Risiken, Zielkonflikten und Unsicherheiten.

Software Engineering wird damit zu einer organisationalen Schlüsselkompetenz. Es stellt Methoden, Modelle, Prozesse und Qualitätssicherungsmechanismen bereit, mit denen Organisationen ihre digitalen Systeme kontrolliert entwickeln, verändern und weiterentwickeln können. Im agentischen Software Engineering umfasst diese Fähigkeit nicht nur die technische Beherrschung von KI-Agenten, sondern zunehmend auch die Fähigkeit, deren Ergebnisse unabhängig zu beurteilen und ihre Verwendung verantwortbar zu entscheiden. Digitale Veränderungsfähigkeit bedeutet damit nicht, jede neue Technologie einsetzen zu können, sondern die Fähigkeit zu besitzen, die eigenen digitalen Produktionsbedingungen gezielt zu verändern und dabei ihre Folgen kontrollieren zu können.

Menschliche Urteilskraft wird damit zu einem Bestandteil der technischen und organisatorischen Vertrauensinfrastruktur. Menschen müssen dabei nicht jede Aktion eines KI-Agenten einzeln kontrollieren~\cite{zhu2026designing}; sie müssen jedoch in der Lage bleiben, dessen Handlungsraum festzulegen, Ergebnisse anhand unabhängiger Evidenz zu beurteilen, Zielkonflikte zu entscheiden und Verantwortung für die Verwendung der Ergebnisse zu übernehmen. Dies erweitert zugleich das Verständnis von V\&V: Für agentisches Software Engineering reicht es nicht aus, ausschließlich das Endprodukt zu prüfen. Auch relevante Entwicklungsprozesse, Agenteninteraktionen, verwendete Wissensquellen, Werkzeuge, Entscheidungs- und Übergabepunkte müssen – soweit für die jeweilige Fragestellung erforderlich – beobachtbar und evaluierbar sein. Vertrauenswürdigkeit entsteht damit nicht durch die bloße Behauptung, eine KI-generierte Software sei leistungsfähig, sondern durch eine nachvollziehbare Vertrauenskette aus Spezifikation, Nachvollziehbarkeit, unabhängiger Evidenz, Tests, formalen oder analytischen Nachweisen und menschlicher Verantwortungsübernahme. Empirische Untersuchungen des Einsatzes von Software-Agenten zeigen bereits, dass menschliche Aufsicht dabei unterschiedliche Formen annimmt – von der Festlegung von Handlungsräumen und gemeinsamer Planung über Laufzeitüberwachung bis zur nachträglichen Prüfung~\cite{dhanorkar2026human}.

Die zentrale Ressource der digitalen Selbsttransformation ist daher nicht allein Rechenleistung, sondern die Fähigkeit, zwischen erzeugbarer Information, technischer Evidenz und begründeter Geltung zu unterscheiden. Diese Fähigkeit muss sowohl individuell als Kompetenz der Software-Ingenieurinnen und -Ingenieure als auch institutionell in Form geeigneter Prozesse, Rollen, Prüfmechanismen und technischer Infrastrukturen verankert werden.

\subsection{Europäische Souveränität}

Aus den technologischen, ökonomischen und epistemischen Pfadabhängigkeiten folgt eine zentrale europäische Gestaltungsaufgabe. Der Begriff der \emph{digitalen Souveränität} ist dabei selbst keine analytisch scharf umrissene Kategorie, sondern ein in Forschung und Politik kontrovers diskutiertes und heterogen gefülltes Konzept~\cite{broeders2023search,fratini2024digital}. Er changiert zwischen einem deskriptiven Verständnis als Fähigkeit zur Kontrolle über kritische digitale Infrastrukturen und einem normativen Verständnis als politisches Leitbild europäischer Selbstbehauptung~\cite{floridi2020fight,santaniello2025attributes}. Digitale Souveränität sollte daher nicht als vollständige Autarkie missverstanden werden. Sie bezeichnet vielmehr die Fähigkeit, kritische digitale Systeme und ihre Produktionsbedingungen verstehen, bewerten, gestalten und bei Bedarf verändern zu können.

Diese Fähigkeit ist nicht voraussetzungslos gegeben. Vergleichende Analysen der europäischen Digitalpolitik zeigen, dass dem politischen Souveränitätsdiskurs nicht in allen Politikfeldern ein entsprechender materieller Politikwandel folgt und dass insbesondere die europäische KI-Politik stärker durch wettbewerbspolitische Zielsetzungen im internationalen Systemvergleich als durch eine grundsätzliche Alternative zur globalen Konkurrenzlogik geprägt ist~\cite{falkner2024digital}. Auch der nationale Souveränitätsdiskurs selbst ist nicht homogen, sondern Ausdruck unterschiedlicher politischer Interessenlagen und Deutungsmuster~\cite{lambach2022narratives}. Eine wissenschaftlich fundierte Auseinandersetzung mit europäischer Souveränität in der Softwareproduktion muss diese Ambivalenz mitreflektieren: Souveränitätspolitik kann Abhängigkeiten wirksam reduzieren, sie kann aber auch als politisches Legitimationsnarrativ fungieren, ohne dass ihr entsprechende strukturelle Investitionen folgen.

Für die digitale Selbsttransformation bedeutet der hier vertretene, produktionsorientierte Souveränitätsbegriff eine Verschiebung des Fokus von fertigen Anwendungen auf die vorgelagerten Produktionsbedingungen digitaler Systeme. Es genügt nicht, europäische Rechenkapazität, Cloud-Infrastruktur oder KI-Modelle aufzubauen. Europa muss auch die Fähigkeit besitzen, die darauf aufsetzende Softwareproduktion zu beherrschen. In Anknüpfung an die in Abschnitt~\ref{de:sec:Gestaltung} eingeführte Unterscheidung von System-, Infrastruktur- und Softwareproduktions-Lock-in lassen sich die zentralen Handlungsfelder entsprechend den drei Ebenen ordnen, auf denen Abhängigkeiten entstehen können:

\begin{itemize}
\item auf der \textbf{Infrastrukturebene}: leistungsfähige und interoperable Entwicklungs- und KI-Agenteninfrastrukturen sowie sichere und auditierbare Engineering- und Agentenplattformen;
\item auf der \textbf{Systemebene}: offene Standards und Schnittstellen für agentisches SE sowie offene digitale Gemeingüter und Open-Source-Ökosysteme;
\item auf der \textbf{Softwareproduktionsebene} im engeren Sinne: hochwertige Daten- und Wissensbestände für SE, europäische Evaluations-, Test- und V\&V-Infrastrukturen, Forschung zu alternativen KI-, Agenten- und Rechnerarchitekturen sowie vor allem die Ausbildung und kontinuierliche Weiterentwicklung entsprechender Engineering-Kompetenzen.
\end{itemize}

Dass Abhängigkeiten im KI-Zeitalter grundsätzlich über klassische Portabilitätsprobleme hinausreichen, zeigt sich auch jenseits des europäischen Kontexts: Aktuelle Analysen zur Beschaffung von KI-Systemen weisen darauf hin, dass KI-spezifische Lock-in-Mechanismen nicht mehr allein auf Softwareschnittstellen beschränkt bleiben, sondern zunehmend geistiges Eigentum und algorithmische Kompetenzen betreffen, da auf proprietären Plattformen trainierte KI-Modelle mitunter untrennbar mit ihrer Trainingsumgebung verbunden sind~\cite{itea2026vendorlockin}. Dies bestätigt aus einer anderen Perspektive die hier vertretene These, dass agentische KI Lock-in tendenziell eine Produktionsebene nach vorne verschiebt.

Die Gesellschaft für Informatik ordnet KI-basiertes SE entsprechend als Schlüsseltechnologie für Deutschlands technologische Innovationskraft und digitale Souveränität ein~\cite{gi2026aikbse}. Diese Einordnung erhält seit Juni 2026 auch eine unmittelbare politische Entsprechung: Mit dem \emph{European Technological Sovereignty Package} hat die Europäische Kommission erstmals einen Politikrahmen vorgelegt, der Halbleiter, Cloud- und KI-Infrastruktur sowie Software explizit als zusammenhängende, sich wechselseitig verstärkende Elemente einer gemeinsamen Souveränitätsstrategie behandelt~\cite{eucom2026techsovereignty}. Bemerkenswert ist dabei, dass die begleitende EU Open Source Strategy Open Source erstmals auf höchster politischer Ebene nicht nur als technisches Werkzeug, sondern als strategisches Instrument europäischer Souveränität verankert -- ein deutlicher Beleg für die praktische Relevanz der oben skizzierten Systemebene. Auch auf zwischenstaatlicher Ebene hat sich diese Priorisierung verfestigt, etwa im Rahmen des deutsch-französischen Gipfels zur europäischen digitalen Souveränität im November 2025~\cite{eucouncil2025summit}. Ergänzend hierzu schlagen aktuelle politiknahe Studien mit dem Konzept eines europäischen EuroStack ein mehrschichtiges Referenzmodell vor, das Rechenzentren, Cloud, Daten, KI-Modelle und Anwendungen als gemeinsam zu gestaltende Souveränitätsarchitektur begreift~\cite{bria2025eurostack}.

Für Deutschland und Europa folgt daraus ein strategischer Perspektivwechsel. Nicht allein die Entwicklung einzelner KI-Modelle sollte gefördert werden, sondern ein leistungsfähiges europäisches Software-Engineering-Ökosystem für das KI-Zeitalter. Forschung, Industrie, öffentliche Hand und Hochschulen benötigen gemeinsame Infrastrukturen, in denen agentische Entwicklungsverfahren entwickelt, erprobt, evaluiert, zertifiziert und in reale Anwendungen überführt werden können. Ob die gegenwärtige politische Dynamik tatsächlich zu einer nachhaltigen Reduktion struktureller Abhängigkeiten führt oder überwiegend deklaratorisch bleibt, ist dabei -- der oben referierten Kritik entsprechend -- eine offene empirische Frage.

Die entscheidende Gestaltungsoption besteht somit darin, die gegenwärtige Phase hoher technologischer Dynamik in Forschung~\cite{SRIA2026_Matrix,abrahao2025software,ahmed2025artificial,hassan2026agentic}, Wirtschaft und Gesellschaft zu nutzen, bevor sich neue Lock-ins verfestigen. Europa muss nicht jeden technologischen Pfad selbst entwickeln. Es muss jedoch die Fähigkeit bewahren, zwischen Pfaden zu wählen, eigene Alternativen hervorzubringen und kritische Abhängigkeiten zu kontrollieren.

\section{Zusammenfassung und Ausblick}
\label{de:sec:Zusammenfassung}

Auch wenn die vierte Digitalisierungsphase nicht abgeschlossen und eine gegenwärtig entstehende Konstellation ist, sind bereits Merkmale und Strukturbrüche erkennbar: 
Die vierte Digitalisierungsphase unterscheidet sich qualitativ von den vorhergehenden Phasen. Digitalisierung transformiert nicht mehr nur Informationen, Prozesse, Organisationen und Wertschöpfungsmodelle. Sie beginnt, ihre eigene Produktionsbasis zu transformieren. Digitale Systeme werden zunehmend mit Hilfe digitaler Systeme entwickelt, getestet, betrieben und weiterentwickelt. Digitalisierung wird damit rekursiv: Sie wird zur digitalen Selbsttransformation.

Der zentrale technische Mechanismus dieser Entwicklung ist der Übergang von generativer KI zur agentischen Softwareproduktion. KI-Systeme erzeugen nicht mehr nur einzelne Codefragmente, sondern können zunehmend zusammenhängende Engineering-Aufgaben verfolgen, Entwicklungsumgebungen bedienen, Software verändern, Tests ausführen und Ergebnisse iterativ verbessern. Damit verschiebt sich die Grenze der Automatisierung von der Codeerzeugung hin zu Teilen des Software-Engineering-Prozesses.

Diese Entwicklung verändert die Arbeitsteilung zwischen Mensch und Maschine, aber sie macht SE nicht obsolet. Im Gegenteil: Je stärker die Implementierung automatisiert wird, desto wichtiger werden die Aufgaben, die den Gegenstand, die Qualität und die Grenzen des Systems bestimmen. Systemintention, Architektur, Qualitätsziele, V\&V, Sicherheit, Nachvollziehbarkeit und Verantwortung werden zu den zentralen Gegenständen ingenieurmäßiger Gestaltung.

Die ökonomische Konsequenz ist eine neue Knappheitsstruktur. Die Grenzkosten bestimmter Formen der Softwareerzeugung sinken, wodurch der Raum wirtschaftlich realisierbarer Software erheblich wachsen kann. Gleichzeitig bleiben Systemverständnis, belastbare Anforderungen, hochwertige Daten, Rechen- und Energieinfrastruktur, unabhängige Evidenz und menschliche Urteilskraft knapp. Zudem realisieren sich die potentiellen Produktivitätsgewinne mit KI nicht automatisch durch die Einführung einzelner Werkzeuge, sondern durch komplementäre Investitionen in Organisation, Prozesse, Kompetenzen und Vertrauensinfrastrukturen.

Damit wird agentisches SE zur Schlüsseltechnologie der vierten Digitalisierungsphase. Es ist die Technologie, mit der die digitale Produktionsbasis selbst neu gestaltet wird. Zugleich wird SE zu einer Schlüsselkompetenz für Unternehmen, Staat und Gesellschaft: Wer die Entstehung und Veränderung digitaler Systeme beherrschen kann, verfügt über eine zentrale Voraussetzung für digitale Wertschöpfung, technologische Innovationsfähigkeit und digitale Souveränität.

Für Deutschland und Europa ist dies von besonderer Bedeutung. Die entscheidende strategische Frage lautet nicht, ob Europa jedes Basismodell oder jede Plattform selbst entwickeln kann. Entscheidend ist, ob Europa die Fähigkeit besitzt, digitale Systeme unabhängig genug zu verstehen, zu gestalten, zu prüfen und weiterzuentwickeln, um kritische Abhängigkeiten kontrollieren und eigene Entwicklungspfade eröffnen zu können.

Die dafür erforderlichen Kompetenzen reichen weit über Prompting oder die Nutzung von KI-Coding-Assistenten hinaus. Software Engineers der nächsten Generation müssen in der Lage sein, komplexe soziotechnische Systeme zu modellieren, Anforderungen zu präzisieren, Architekturen zu entwerfen, KI-Agenten zu orchestrieren, Evidenz für die Qualität ihrer Ergebnisse zu erzeugen und die Grenzen maschineller Entscheidungen zu erkennen. Sie werden damit weniger zu manuellen Produzenten von Code als zu Gestaltern und Managern digitaler Produktionssysteme.

Für die künftige Forschung ergeben sich daraus die folgenden Desiderate: Die SE-Forschung muss Modelle für Mensch-Agenten-Kollaboration, agentische Entwicklungsprozesse, Multi-Agenten-Systeme, maschineninterpretierbare Spezifikationen, kontinuierliche V\&V und Vertrauensinfrastrukturen entwickeln. Die Ausbildung muss diese Fähigkeiten systematisch vermitteln und dabei ein solides Fundament in Informatik und klassischem SE mit neuen Kompetenzen für agentische Systeme verbinden.

Gleichzeitig muss die Politik die Entstehung eines europäischen Engineering"=Ökosystems unterstützen. Dazu gehören offene Standards und digitale Gemeingüter, sichere und interoperable Agenteninfrastrukturen, europäische Forschungs- und Testinfrastrukturen sowie langfristig angelegte Programme für Forschung, Ausbildung und Transfer.

Die zentrale Frage der vierten Digitalisierungsphase lautet daher nicht, wie viel Software KI erzeugen kann. Sie lautet vielmehr, ob Menschen, Organisationen und Gesellschaften die Fähigkeit bewahren und ausbauen, gezielt die Entstehung digitaler Systeme zu bestimmen. Genau darin liegt die strategische Bedeutung des SE.

Wenn Digitalisierung in eine Phase der digitalen Selbsttransformation eintritt, wird SE von einer wichtigen Ingenieurdisziplin zu einer Voraussetzung dafür, die digitale Transformation selbst gestalten zu können. Agentisches SE geht damit über eine neue Form der Softwareentwicklung hinaus: Es ist eine der zentralen soziotechnischen Voraussetzungen digitaler Souveränität Deutschlands und Europas. Ob und wie weit diese Möglichkeiten tatsächlich in strukturelle Unabhängigkeit übersetzt werden, hängt jedoch von institutionellen und politischen Rahmenbedingungen ab.

\reprintbibliography

\end{document}